\documentclass[%
 aip,
 amsmath,amssymb,
 reprint,%
]{revtex4-2}

\usepackage{graphicx}
\usepackage{dcolumn}
\usepackage{bm}

\usepackage[utf8]{inputenc}
\usepackage[T1]{fontenc}
\usepackage{mathptmx}
\usepackage{etoolbox}
\usepackage{hyperref}
\hypersetup{
    colorlinks=true,
    linkcolor=blue,
    filecolor=magenta,      
    urlcolor=cyan,
    }
\makeatletter 
\let\@fnsymbol\@fnsymbol@latex \@booleanfalse\altaffilletter@sw 

\def\doauthor#1#2#3{%
  \ignorespaces#1\unskip\@listcomma
  \begingroup
   #3%
  \@if@empty{#2}{\endgroup{}{}}{\endgroup{\comma@space}{}*}%
  \space \@listand 
}%
\makeatother

\makeatletter
\def\@email#1#2{%
 \endgroup
 \patchcmd{\titleblock@produce}
  {\frontmatter@RRAPformat}
  {\frontmatter@RRAPformat{\produce@RRAP{*#1\href{mailto:#2}{#2}}}\frontmatter@RRAPformat}
  {}{}
}%
\makeatother

\DeclareMathAlphabet{\mathcal}{OMS}{cmsy}{n}{n}

\begin{document}

\preprint{AIP/123-QED}

\title{Pound-Drever-Hall locking scheme free from Trojan operating points}
\author{Manuel Zeyen}
\email[\textbf{Authors to whom correspondence should be addressed: }]{zeyenm@phys.ethz.ch; aldo.antognini@psi.ch}
\author{Lukas Affolter}
\affiliation{ 
Institute for Particle Physics and Astrophysics, ETH, 8093 Zurich, Switzerland.
}
\author{Marwan Abdou Ahmed}
\affiliation{Institut für Strahlwerkzeuge, Universität Stuttgart, Pfaffenwaldring 43, 70569 Stuttgart, Deutschland.}

\author{Thomas Graf}
\affiliation{Institut für Strahlwerkzeuge, Universität Stuttgart, Pfaffenwaldring 43, 70569 Stuttgart, Deutschland.}

\author{Oguzhan Kara}
\affiliation{ 
Institute for Particle Physics and Astrophysics, ETH, 8093 Zurich, Switzerland.
}
\author{Klaus Kirch}
\author{Miroslaw Marszalek}
\affiliation{ 
Institute for Particle Physics and Astrophysics, ETH, 8093 Zurich, Switzerland.
}%
\affiliation{ 
Paul Scherrer Institute, 5232 Villigen, Switzerland.
}%
\author{François Nez}
\affiliation{Laboratoire Kastler Brossel, Sorbonne Université, CNRS, ENS-Université PSL, Collège de France, 75252 Paris Cedex 05, France.
}
\author{Ahmed Ouf}
\author{Randolf Pohl}
\author{Siddharth Rajamohanan}
\affiliation{ 
Johannes Gutenberg-Universität Mainz, QUANTUM, Institut für Physik \& Exzellenzcluster PRISMA, 55128 Mainz, Germany.
}

\author{Pauline Yzombard}
\affiliation{Laboratoire Kastler Brossel, Sorbonne Université, CNRS, ENS-Université PSL, Collège de France, 75252 Paris Cedex 05, France.
}

\author{Aldo Antognini}\affiliation{ 
Institute for Particle Physics and Astrophysics, ETH, 8093 Zurich, Switzerland.
}
\affiliation{ 
Paul Scherrer Institute, 5232 Villigen, Switzerland.
}
\author{Karsten Schuhmann}
\affiliation{ 
Institute for Particle Physics and Astrophysics, ETH, 8093 Zurich, Switzerland.
}

\date{\today}

\begin{abstract}
The Pound-Drever-Hall (PDH) technique is a popular method for stabilizing the frequency of a laser to a stable optical resonator or, vice versa, the length of a resonator to the frequency of a stable laser. We propose a refinement of the technique yielding an "infinite"  dynamic (capture) range so that a resonator is correctly locked to the seed frequency, even after large perturbations. The stable but off-resonant lock points (also called Trojan operating points), present in conventional PDH error signals, are removed by phase modulating the seed laser at a frequency corresponding to half the free spectral range of the resonator. We verify the robustness of our scheme experimentally by realizing an injection-seeded Yb:YAG thin-disk laser. We also give an analytical formulation of the PDH error signal for arbitrary modulation frequencies and discuss the parameter range for which our PDH locking scheme guarantees correct locking. Our scheme is simple as it does not require additional electronics apart from the standard PDH setup and is particularly suited to realize injection-seeded lasers and injection-seeded optical parametric oscillators.

\end{abstract}

\maketitle

\section{Introduction}\label{sec:intro}
The Pound-Drever-Hall (PDH) locking technique  \cite{drever_laser_1983} is a popular method to stabilize the frequency of a laser to a stable optical resonator. It has been used to achieve lasers with sub-Hertz linewidth \cite{kessler_sub-40-mhz-linewidth_2012,schmid_simple_2019} and is applied in a wide range of fields, such as gravitational wave detection \cite{willke_stabilized_2010}, atomic physics \cite{hond_medium-finesse_2017,legaie_sub-kilohertz_2018} and metrology, \cite{grinin_two-photon_2020} just to name a few. The PDH method can also be used in the opposite way to stabilize the length of an optical resonator to a stable single-frequency laser with equally numerous applications \cite{wulfmeyer_2-m_2000,chen_injection-seeded_2022,zhang_seeded_2020,ricciardi_sub-kilohertz_2015}. 

Despite its widespread and long-standing application, this technique is continuously refined and adapted to specific applications \cite{cygan_active_2011,gatti_wide-bandwidth_2015,izumi_self_nodate,zeng_stabilizing_2021,wang_artificial_2019}. The dynamic range of the standard PDH technique is limited by the additional zero crossings of the error signal at off-resonant frequencies. This typically limits the dynamic range (or capture range) of the lock to a fraction of the free spectral range (FSR) of the resonator given by
\begin{equation}
    \Delta f = \frac{c}{2L},
\end{equation}
where $c$ is the speed of light in the resonator and $L$ is the length of the (linear) resonator. The PDH error signal has a zero crossing right in the middle between two adjacent resonator modes at frequency detuning $\nu = \Delta f/2$ from the resonator mode. This zero crossing represents a stable lock point, where laser and resonator are stabilized in a totally off-resonant state. Such undesired stable operating points are also called Trojan operating points \cite{homer_odyssey_nodate, yen-ting_wang_practical_2013}. Since the modulation frequency is typically much smaller than the FSR of the resonator, a large disturbance (causing large laser frequency or resonator length variations) may lead to an erroneous stabilization on the Trojan operating point. When this occurs, the correct lock point must be restored either manually or via an automated process, which requires dedicated electronics and can take up to several ms \cite{schutte_experimental_2016,haze_note_2013,guo_automatic_2022}.

In this paper, we demonstrate a simple way to avoid off-resonant Trojan operating points in a PDH error signal, by modulating the seed laser at $\nu_{M} = \Delta f/2$, i.e. at half the FSR of the resonator. In doing so, the off-resonant lock point between two resonances is made unstable, resulting in an ``infinite'' capture range so that re-locking always occurs on a resonance independently of the size of the perturbation. Our scheme is particularly well suited for injection-seeding lasers or optical parametric oscillators.

The paper is organized as follows: In section \ref{sec:Theory} we present the theory of the PDH error signal, highlighting the peculiarities related to the use of $\nu_{M}=\Delta f/2$. We also emphasize the parameter range in which the locking scheme works best and we link our scheme to recent ideas which extend the linear range of the PDH error signal\cite{evans_lock_2002,li_broadening_2021,miyoki_expansion_2010}. In section \ref{sec:Experimental_Results} we present an implementation of our scheme in an injection-seeded thin-disk laser (TDL).

\section{PDH scheme without Trojan operating points}\label{sec:Theory}

\subsection{Analytical expression for the classic PDH error signal}
In general, a resonator, where the losses mainly occur at the end-mirrors, can be simplified to a two-mirror resonator, where the mirrors have power reflectivities $R_1$ and $R_2$. Without loss of generality, possible intra-resonator gain can be included in $R_2$ so that $R_2 > 1$ is possible.
For such a general resonator we define its finesse \cite{pollnau_spectral_2020} as 
\begin{equation}
    \mathcal{F} = \frac{2\pi}{-\textnormal{ln}\left(R_{1}R_{2}\right)}.
\end{equation}
$\mathcal{F}$ is a measure of the sharpness of the resonances and can be approximated as the ratio
\begin{equation}
    \mathcal{F}\approx \frac{\Delta f}{\delta f},
\end{equation}
where $\delta f$ is the FWHM linewidth of the resonances. In this study $\nu$ denotes the relative detuning between the seed laser frequency and the nearest resonance frequency of the resonator $\textnormal{TEM}_{00}$ modes.

Depending on $\nu$, part of the phase modulated seed light is reflected from the resonator. Its electric field amplitude and phase relative to the input light is given by the complex reflection coefficient \cite{siegman_lasers_1986}
\begin{equation}\label{eq:F(nu)}
    F(\nu )=\sqrt{R_{1}}-\frac{(1-R_{1})\sqrt{R_{2}}\exp{\left[2i(\pi \frac{\nu}{\Delta f}+\phi)\right]}}{1-\sqrt{R_{1}R_{2}}\exp{\left[2i(\pi \frac{\nu}{\Delta f}+\phi)\right]}},
\end{equation}
where $\phi$ is an additional phase shift which the light might acquire over one resonator round trip (e.g. by propagating through a gain medium). For simplicity, in the following we set $\phi=0$.

From this reflection coefficient, the well known PDH error signal is obtained as \cite{black_introduction_2001}
\begin{eqnarray}\label{eq:e(nu)}
    \epsilon(\nu,\nu_{M})=&&-2\sqrt{P_{c}P{s}}\ \textnormal{Im}\left[F(\nu )F^{*}(\nu + \nu_{M})\right. \nonumber\\&&\left. -F^{*}(\nu )F(\nu - \nu_{M})\right],
\end{eqnarray}
where $\nu_{M}$ is the modulation frequency, $P_{c}$ and $P_{s}$ are the power in the carrier and the sidebands of the seed laser respectively, Im[...] takes the imaginary part and * denotes complex conjugation. Note that this is the error signal only for the case of demodulation at $\nu_{M}$ and phase delay $\Delta \varphi = 0$ (see Subsec.\ref{subsec:demod_phase}). 

\begin{figure}
\includegraphics[width=0.95\linewidth]{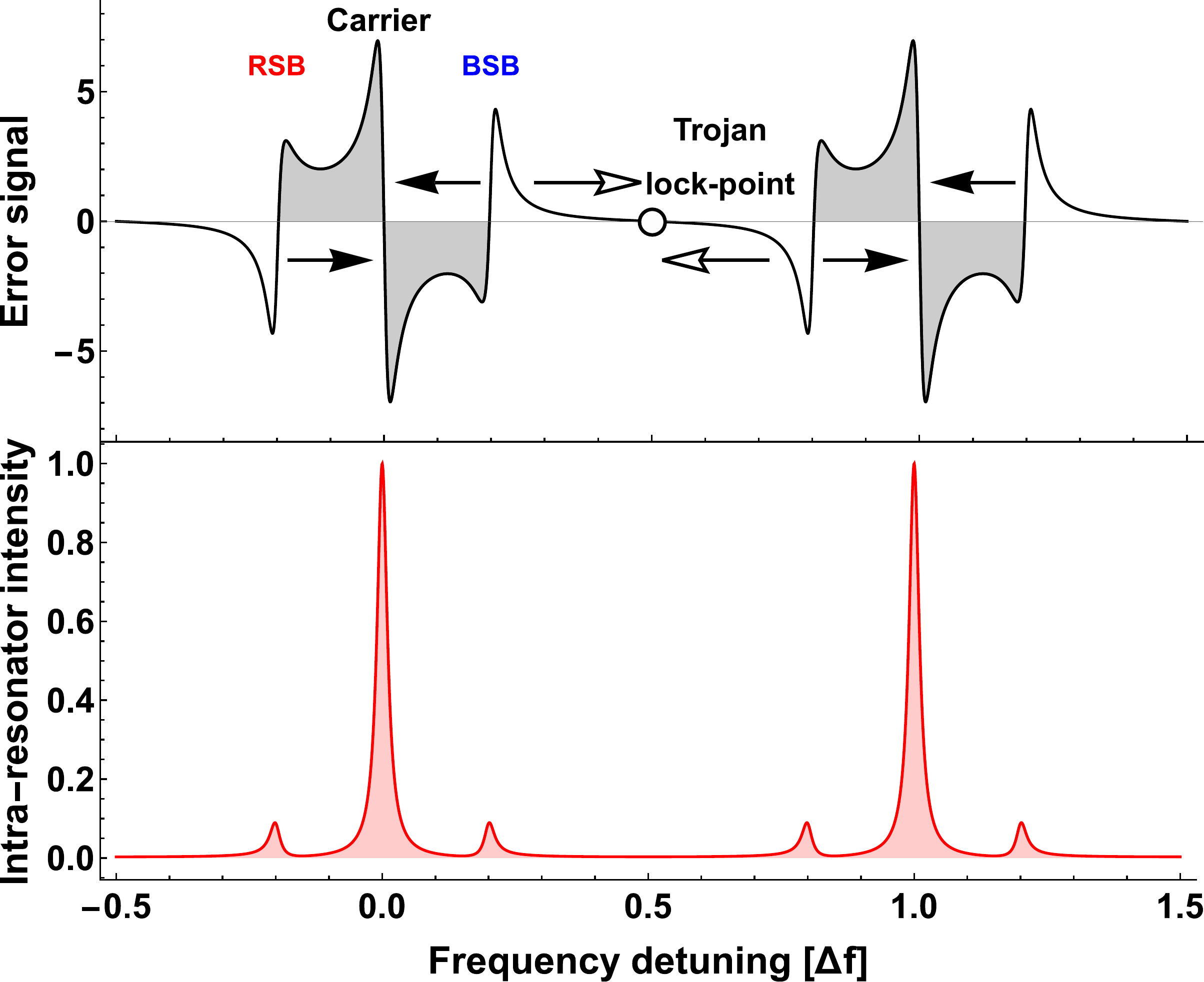}
\caption{\textbf{Top:} Simulation of a typical open-loop PDH error signal obtained by scanning the resonator length. The shaded area indicates the capture range of the feedback loop, i.e. the detuning region between resonator mode and seed laser frequencies that can be corrected. The arrows indicate the direction of the correction of the feedback loop. In this example the laser was modulated at $\nu_{M} = 0.2 \Delta f$ so that the red and blue sideband (RSB and BSB, respectively) lie at $\pm 0.2\Delta f$ from the carrier frequency. \textbf{Bottom:} Corresponding simulated intra-resonator intensity.}
\label{fig:BasicPDH}
\end{figure}

The modulation frequency $\nu_{M}$ can be expressed in terms of the FSR as $\nu_{M} = \xi \Delta f$, where $0 < \xi \leq 1$ so that the PDH error signal can be re-written as
\begin{eqnarray}\label{eq:e(nu)_xi}
    \epsilon(\nu,\xi)=&&-2\sqrt{P_{c}P{s}}\ \textnormal{Im}\left[F(\nu )F^{*}(\nu + \xi \Delta f)\right. \nonumber\\&&\left. -F^{*}(\nu )F(\nu - \xi \Delta f)\right].
\end{eqnarray}
Inserting Eq.~\eqref{eq:F(nu)} into Eq.~\eqref{eq:e(nu)_xi} we find
\begin{eqnarray}\label{eq:e(nu)_general}
    \epsilon(\nu,\xi) =&&4\sqrt{P_{c}P_{s}} \frac{\sin\left(\xi \pi\right)}{G_{1}(\nu)}\left[\frac{G_{2}(\nu,\xi)}{G_{1}(\nu + \xi \Delta f)}\right. \nonumber\\&&-\left.\frac{G_{2}(\nu,-\xi)}{G_{1}(\nu - \xi \Delta f)}\right],
\end{eqnarray}
with
\begin{equation}
    G_{1}(\nu) = 1+\gamma^2-2\gamma\cos\left(\frac{2\pi\nu}{\Delta f}\right)
\end{equation}
and
\begin{eqnarray}
    G_{2}(\nu,\xi) =&& (\gamma^2 R_{1}-R_{2})\cos\left(\xi\pi\right)\nonumber\\&&+\left[\gamma-\gamma^3+\gamma(R_{2}-R_{1})\right]\cos\left(\frac{2\pi\nu}{\Delta f}+\xi\pi\right),
\end{eqnarray}
where
\begin{equation}
    \gamma = \sqrt{R_{1}R_{2}}.
\end{equation}
Instead of using the parameter $\gamma$, these equations can also be expressed in terms of the finesse by performing the substitution
\begin{equation}
    \gamma = \exp\left(-\frac{\pi}{\mathcal{F}}\right).
\end{equation}
Figure~\ref{fig:BasicPDH} shows an example of an error signal for $\xi=0.2$ and $\mathcal{F} = 40$ ($R_{1}=0.6$ and $R_{2}=1.42$). The capture range of the error signal is indicated as gray shaded area. The black full arrows indicate the direction of the feedback loop correction moving the system to the correct lock point ($\textnormal{TEM}_{00}$ resonance), whereas the open arrows indicate the region where the feedback loop steers the system erroneously towards the Trojan operating point. In the next subsection we present a method to avoid this stable but unwanted lock point.
\subsection{Modulating at half the free spectral range}
\begin{figure}
\centering
\includegraphics[width=0.95\linewidth]{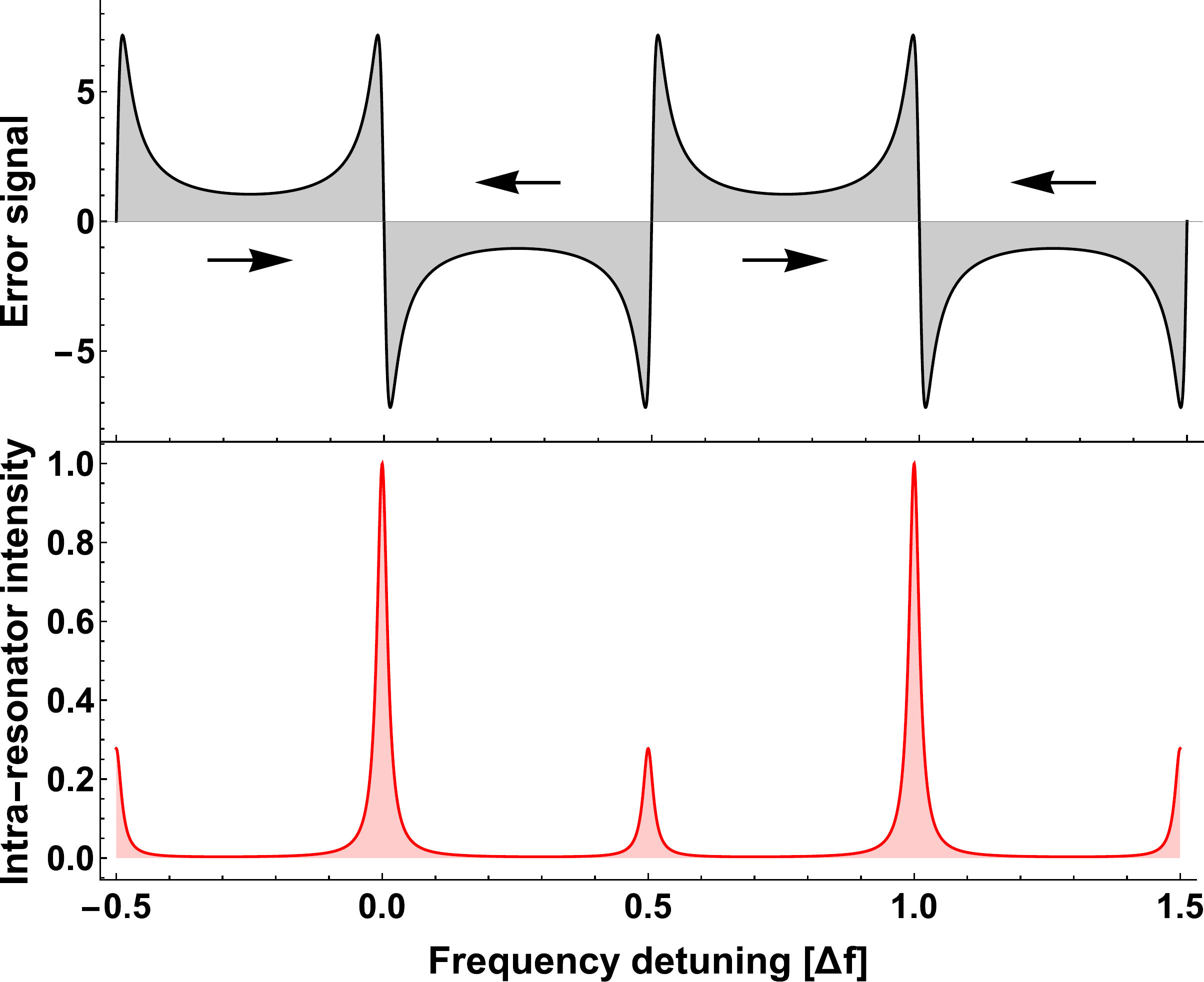}
\caption{Similar to Fig. \ref{fig:BasicPDH} but for $\nu_{M} = \Delta f /2$. The shaded gray area covers the whole free spectral range, so that the capture range of the feedback loop is effectively "infinite". In this case the red and blue sidebands overlap at the detuning $\Delta f /2$.}
\label{fig:OurPDH}
\end{figure}

If we modulate the seed laser at half the FSR of the resonator, i.e. $\nu_{M} = \Delta f/2$ or $\xi = 1/2$, Eq.~\eqref{eq:e(nu)_general} simplifies to
\begin{equation}\label{eq:e(nu,xi=0.5)}
    \epsilon(\nu,\frac{1}{2}) = 8\sqrt{P_{c}P_{s}}\frac{\left[\gamma^3-\gamma(1+R_{2}-R_{1})\right]\sin\left(\frac{2\pi\nu}{\Delta f}\right)}{1+\gamma^4-2\gamma^2\cos\left(\frac{4\pi\nu}{\Delta f}\right)}.
\end{equation}
In this case the blue and the red sidebands from neighboring resonances overlap and the error signal is free from Trojan operating points, as stressed by the shaded area spanning the whole region between two resonances in Fig.~\ref{fig:OurPDH}.

Since $0 < \gamma < 1$, we find 
\begin{equation}
\epsilon(\nu,\frac{1}{2}) \begin{cases}
      \leq 0 & \text{for } \nu \in [-\Delta f/2,0]\\
      > 0 & \text{for } \nu \in (0,\Delta f/2]
      \end{cases} 
\end{equation}
as the denominator in Eq.~\eqref{eq:e(nu,xi=0.5)} is a positive quantity. This guarantees that the only stable lock points coincide with the resonator resonances. The feedback loop will thus always re-lock on the nearest resonance independently of the size of the perturbation.

\subsection{Tolerable mismatch in the case \texorpdfstring{$\nu_{M}\neq\Delta f/2$}{Text}}
\begin{figure}
    \includegraphics[width=0.95\linewidth]{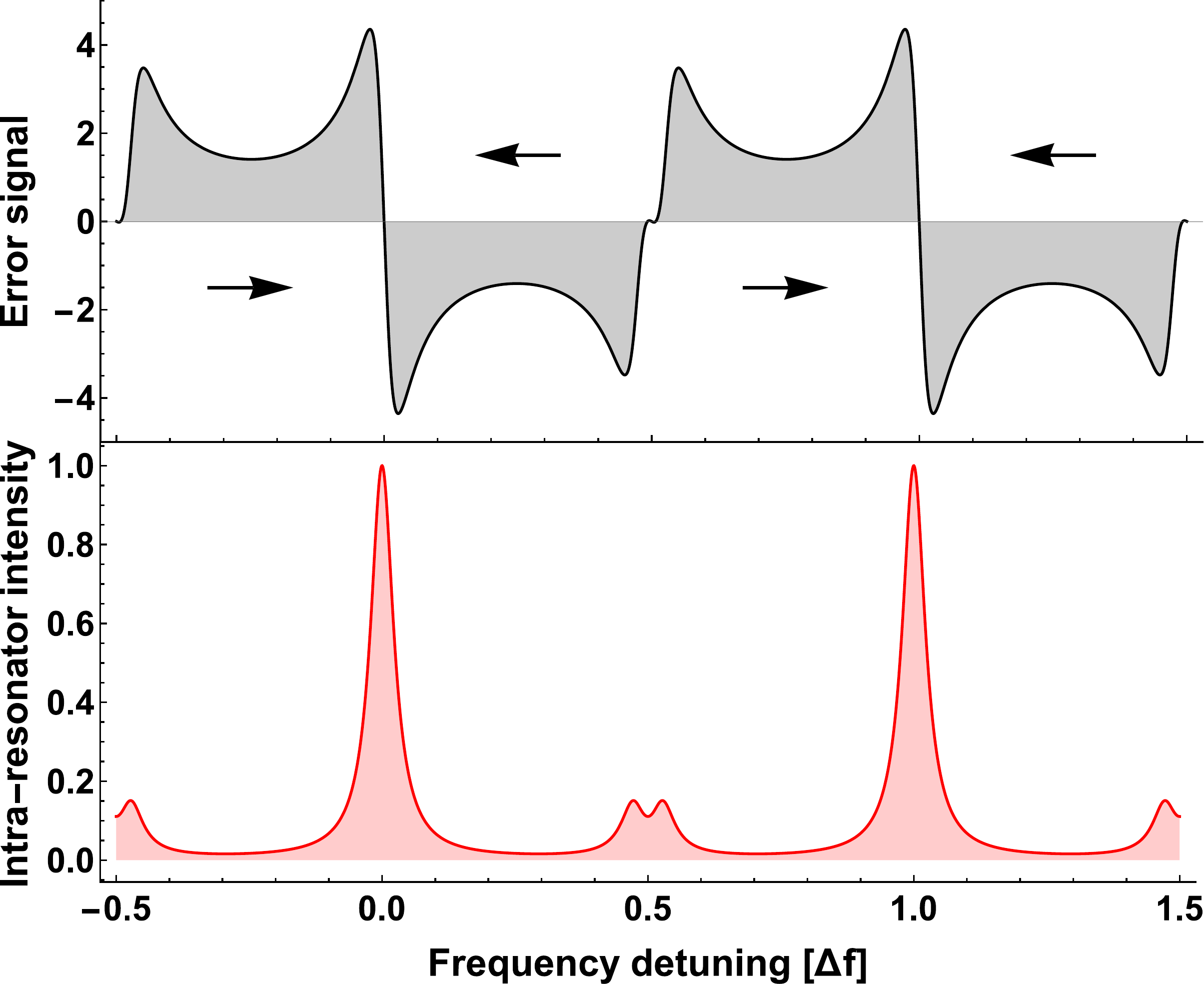}
    \caption{\textbf{Top:} Similar to Fig.~\ref{fig:OurPDH} but in this simulation the seed laser was modulated with $\nu_{M} = \xi_{+}$ so that the error signal has a saddle point at $\Delta f/2$. In this simulation we use $\mathcal{F}=19$ ($\gamma=0.848$) and $\nu_{M} = 0.526 \Delta f$, as obtained from Eq.~\eqref{eq:xi_zeros}. The shaded gray area still covers the whole free spectral range so that the capture range of the feedback loop is still "infinite". \textbf{Bottom:} Corresponding simulated intra-resonator intensity. The red and blue sidebands no longer perfectly overlap at $0.5 \Delta f$, but the error signal is still usable.}
    \label{fig:OurPDH_Mismatch}
\end{figure}

\begin{figure}
\includegraphics[width=0.95\linewidth]{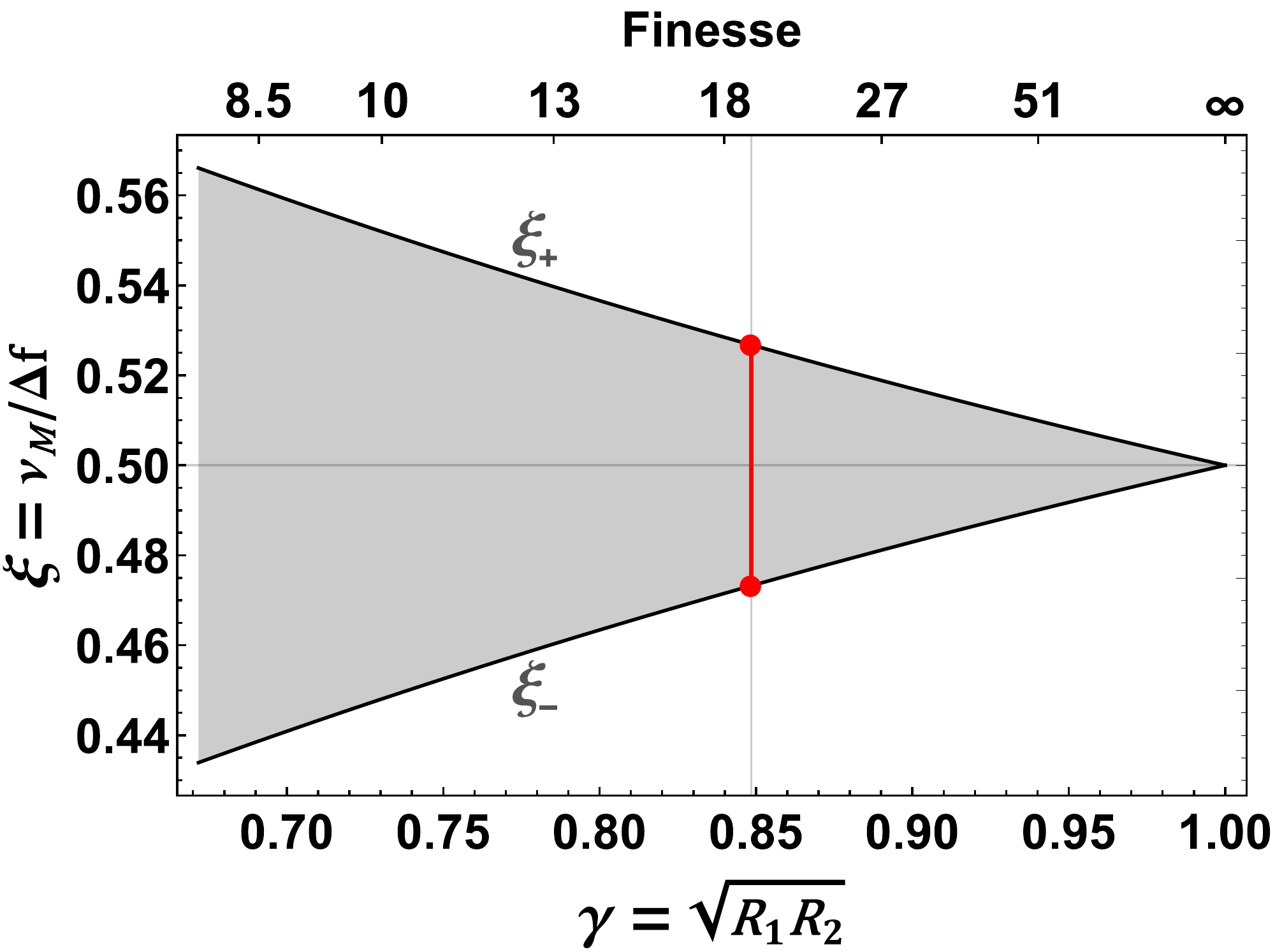}
\caption{$\xi_{-}$ and $\xi_{+}$ versus $\gamma$ (see Eq.~\eqref{eq:xi_zeros}). The gray shaded area $(\xi_{-}<\xi<\xi_{+})$ indicates the allowed window of modulation frequencies in terms of $\Delta f$ yielding an "infinite" capture range (without Trojan operating points). The red line illustrates an example where $\mathcal{F}=19$ (see Sec.~\ref{sec:Experimental_Results}), allowing for $\nu_{M} \in [0.475\Delta f,0.525 \Delta f]$.}
\label{fig:Mismatch}
\end{figure}

 For the practical realization of this locking scheme, it is key to understand the sensitivity of the error signal to a mismatch $\nu_{M} \neq \Delta f/2$. The maximal mismatch allowed to still achieve correct locking over the whole range is reached when the error signal has a saddle point at $\nu = \Delta f/2$ so that the sign of the error signal still only flips at $\nu = \Delta f/2$ (see Fig.~\ref{fig:OurPDH_Mismatch}).
To find the modulation frequency for which the saddle point appears in the error signal, we calculate the derivative of the error signal at $\nu=\Delta f/2$:
\begin{widetext}
\begin{equation}\label{eq:e(nu)_deriv}
    \frac{\partial \epsilon(\nu,\xi)}{\partial \nu}\bigg|_{\nu=\frac{\Delta f}{2}} = 16\sqrt{P_{c}P_{s}}\gamma\pi\frac{2\left[(\gamma^2 R_{1}-R_{2})\cos(2\xi\pi)-\gamma+\gamma^3\right]+\gamma_{0}}{\left(1+\gamma\right)^2\left( 1+\gamma^2+2\gamma \cos(2\xi \pi)\right)^2}\sin(\xi\pi)^2,
\end{equation}
\end{widetext}
where $\gamma_{0} = 1-\gamma^4-2\gamma (R_{2}-R_{1})+(\gamma^2-1)(R_{1}+R_{2})$. The non-trivial zeros of Eq.~\eqref{eq:e(nu)_deriv} are
\begin{equation}\label{eq:xi_zeros}
    \xi_{\pm} = \frac{1}{2}\pm\frac{1}{2\pi}\arccos{\frac{\gamma_{0}-2\gamma(1-\gamma^2)}{2(\gamma^2R_{1}-R_{2})}},
\end{equation}
which correspond to the minimal and maximal values that $\xi$ can take in order to avoid additional zero crossing in the PDH error signal. The allowed mismatch for a general resonator is illustrated in Fig.~\ref{fig:Mismatch}. The figure shows the two solutions obtained in Eq.~\eqref{eq:xi_zeros} versus the gain factor $\gamma$. The gray shaded area between the two curves indicates the allowed region of modulation frequencies in terms of $\xi$, where $\xi = 0.5$ means modulation at exactly half the FSR. Clearly, higher values of $\mathcal{F}$ (or $\gamma$) reduce the tolerance for a mismatch between $\nu_{M}$ and $\Delta f/2$. For this reason our scheme is particularly suited for injection seeding of laser resonators with not too high finesse. In red is indicated a typical value of $\mathcal{F} \approx 19$ (or $\gamma \approx 0.85$, with $R_{1}=0.5$ and $R_{2}=1.44$) which we used in our experimental verification (see Sec.~\ref{sec:Experimental_Results}). In this case we could tolerate $\nu_{M} \in [0.475 \Delta f,0.525 \Delta f]$, i.e. a mismatch between $\nu_{M}$ and $\Delta f$ of about $\pm 5$\,\%. This requirement is easy to satisfy either by mechanical design or by using a frequency-tunable modulator. 

\subsection{Dependence of the locking scheme on resonator finesse}
\begin{figure}
\includegraphics[width=0.95\linewidth]{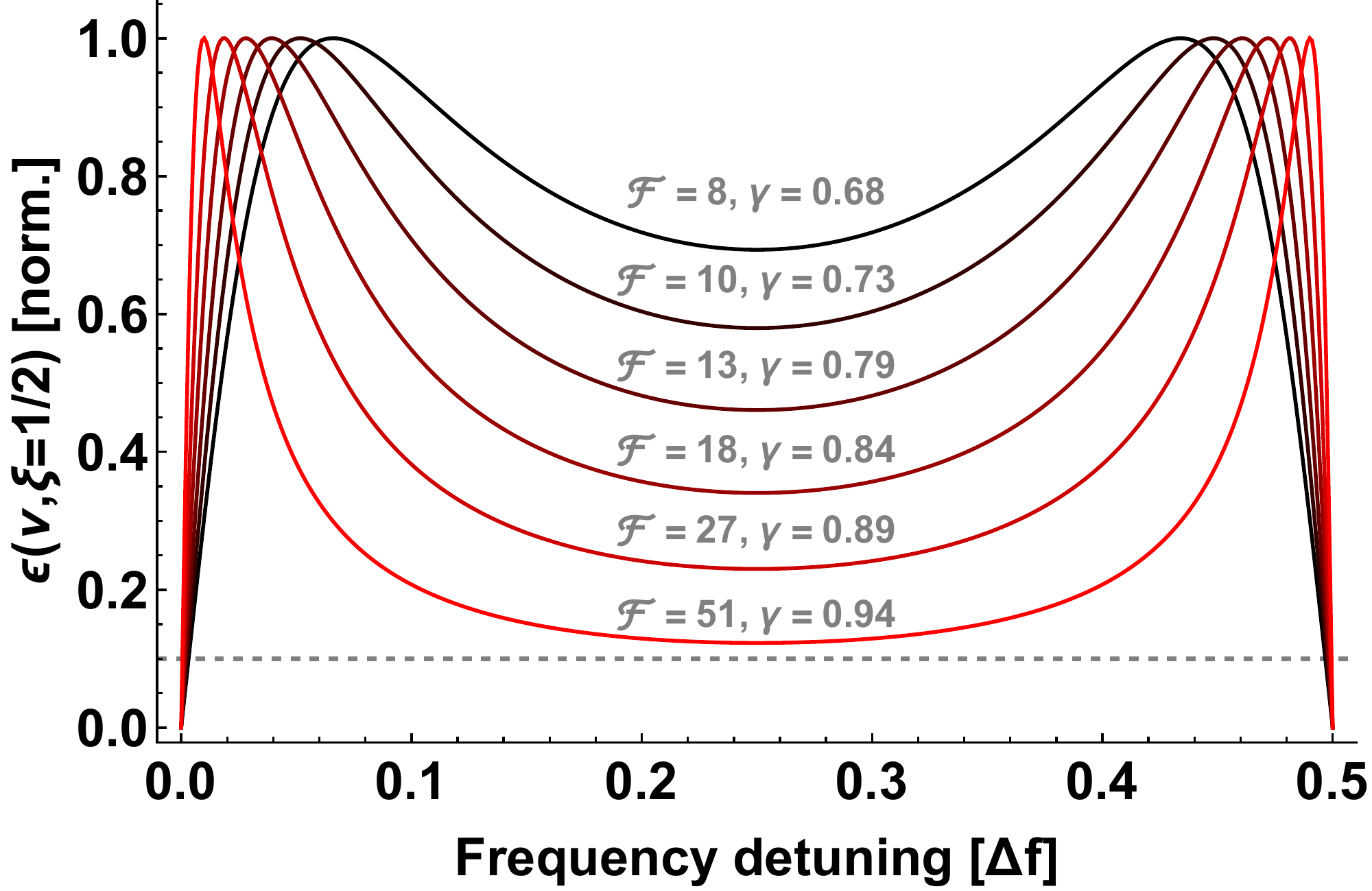}
\caption{Simulated PDH error signal of Eq.~\eqref{eq:e(nu,xi=0.5)} for various values of resonator finesse $\mathcal{F}$ (gain factors $\gamma$). The horizontal dashed line indicates the 10\,\% level of the error signal. The error signals are normalized to have the same maximum.}
\label{fig:OurPDH_CloseZero}
\end{figure}

Our PDH locking scheme with $\nu_{M}\approx\Delta f/2$ is best suited for locking a low-finesse resonator to a seed laser. Indeed, Fig.~\ref{fig:OurPDH_CloseZero} illustrates how the error signal approaches zero between the carrier and the sideband for increasing $\mathcal{F}$. If the electronic noise on the error signal is too large, the error signal can have random zero crossings which would lead to Trojan operating points and jeopardize the whole idea of the half FSR locking. For example, a (conservative) noise level of 10\,\%, as indicated by the dashed horizontal line in Fig.~\ref{fig:OurPDH_CloseZero}, requires the finesse of the seeded resonator to be $\mathcal{F} \lesssim 55$ (or $\gamma \lesssim 0.95$).
\subsection{Dependence of the locking scheme on the demodulation phase shift}\label{subsec:demod_phase}
\begin{figure}
\includegraphics[width=0.95\linewidth]{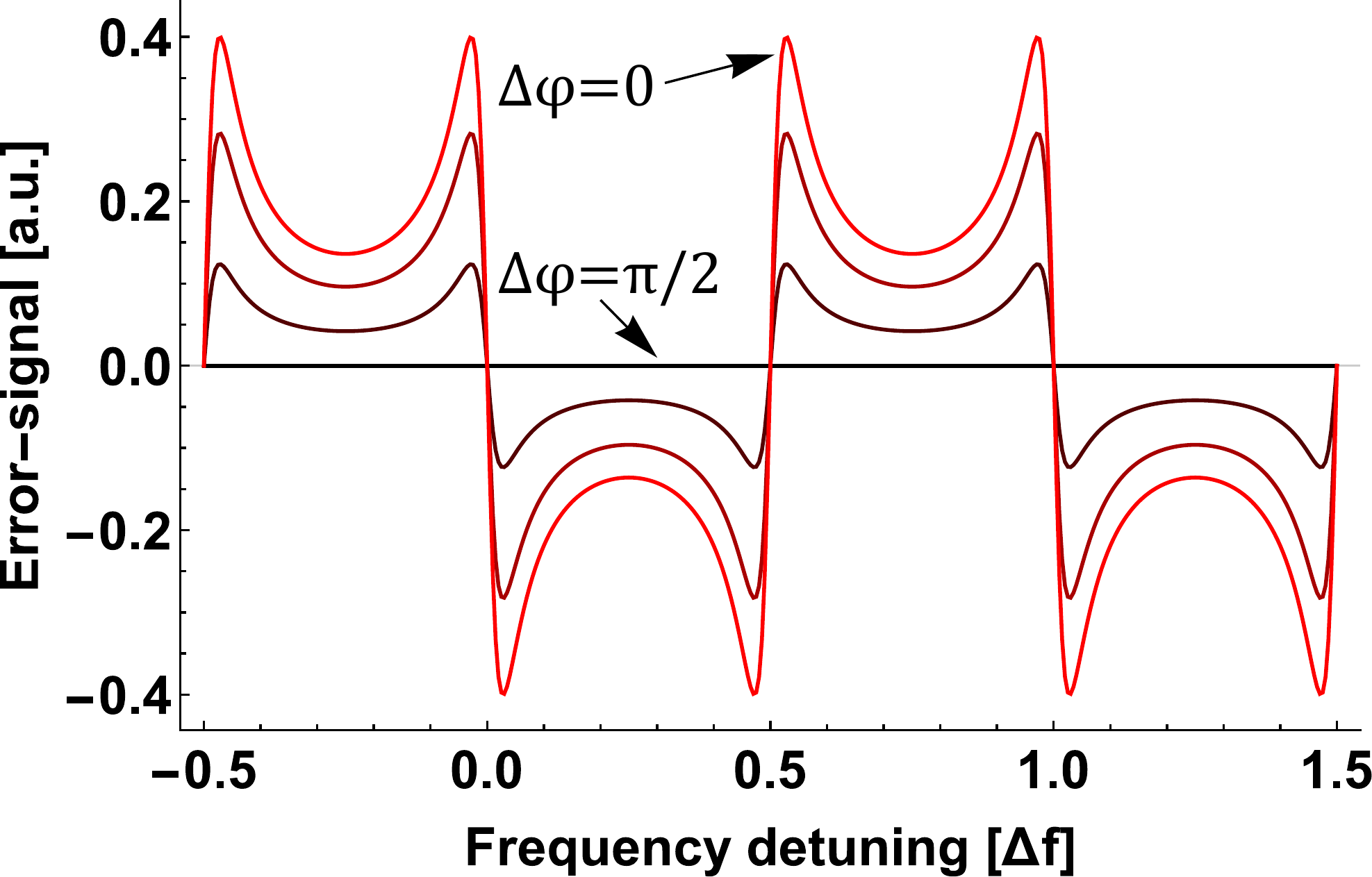}
\caption{Simulated PDH error signal with $\nu_{M} = \Delta f/2$ for various phase shifts $\Delta \phi$ in the demodulation process. The error signal is zero when demodulated in quadrature ($\Delta \phi = \pi/2$). Note that in this simulation the error signals are not all obtained with Eq.~\eqref{eq:e(nu,xi=0.5)} but have to be calculated separately for different values of $\Delta \varphi$.}
\label{fig:OurPDH_phase}
\end{figure}

Strikingly, our scheme is rather insensitive to variations of the phase shift $\Delta \varphi$ used in the demodulation (see Sec.~\ref{sec:Experimental_Results}). While the amplitude of the error signal decreases for $\Delta \varphi \neq 0$, the overall shape stays the same as shown in Fig.~\ref{fig:OurPDH_phase}. This propertiy simplifies the practical implementation and optimization of the PDH loop parameters.
\subsection{Comparison to schemes with a large linear locking range}
\begin{figure}
\includegraphics[width=0.85\linewidth]{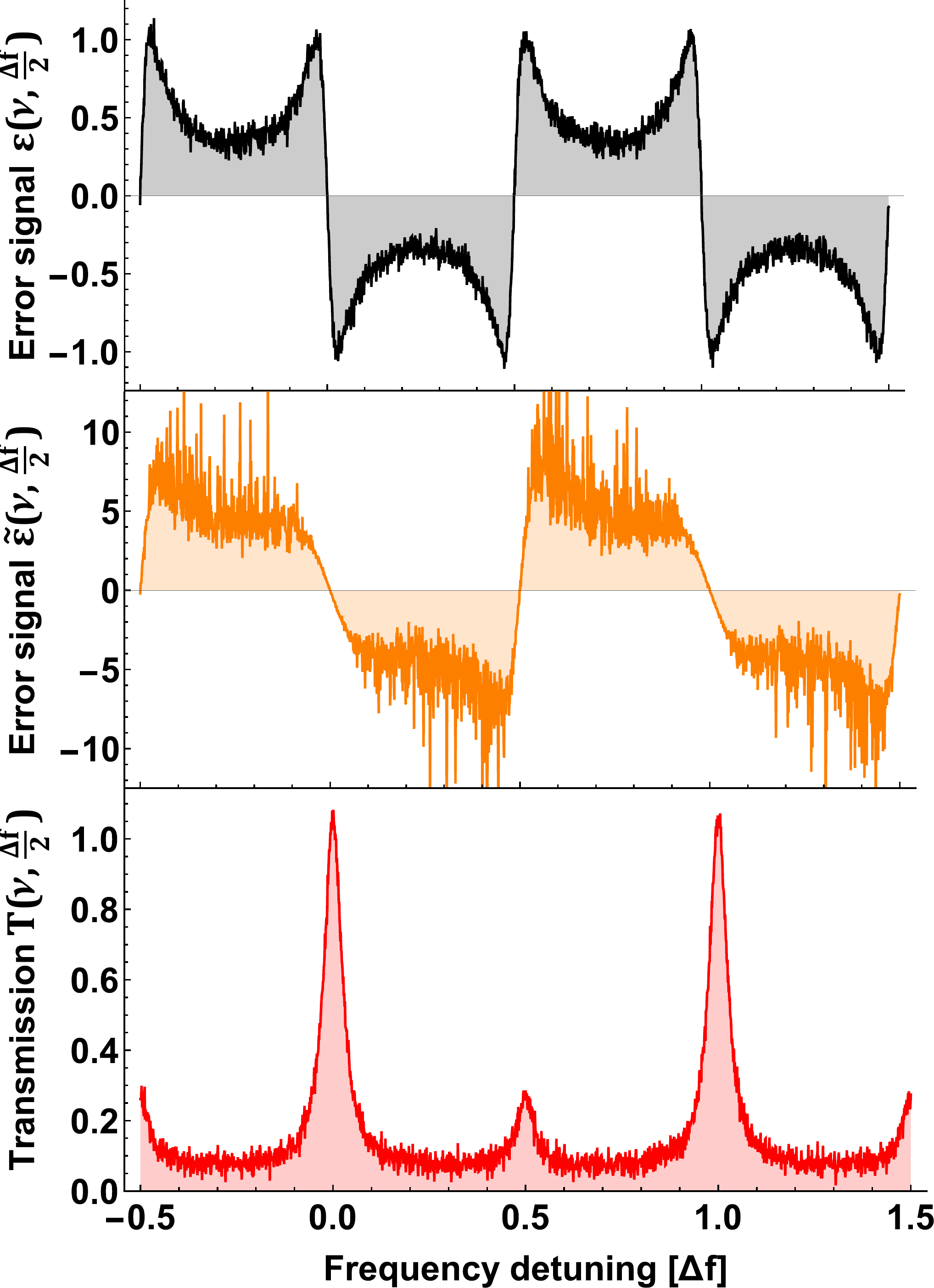}
\caption{\textbf{Top:} Similar to Fig.~\ref{fig:OurPDH} but the simulated error signal contains (Gaussian) noise. \textbf{Middle:} Error signal $\tilde{\epsilon}(\nu)$ obtained with Eq.~\eqref{eq:linearizePDH}. \textbf{Bottom: } Corresponding simulated intra-resonator intensity.}
\label{fig:linearPDH}
\end{figure}

Recently, a PDH scheme has been proposed in which the conventional error signal is divided by the transmitted power $T(\nu,\nu_{M})$ from the resonator \cite{evans_lock_2002,li_broadening_2021} to obtain a new error signal
\begin{equation}\label{eq:linearizePDH}
    \tilde{\epsilon}(\nu,\nu_{M}) = \frac{\epsilon(\nu,\nu_{M})}{T(\nu,\nu_{M})}
\end{equation}
with an increased linear dynamic range. Our scheme can be combined with this approach, resulting in an error signal as shown in Fig.~\ref{fig:linearPDH}. The modified error signal merges the advantage of both schemes: an "infinite" (over the whole FSR) capture-range and a more linear behavior. The new error signal is more noisy between the resonance and the sidebands since $T(\nu,\nu_{M})$ (i.e. the denominator of Eq. \eqref{eq:linearizePDH}) is small in this region. However, for this large detuning the noise should not excessively disturb the proper working of the feedback loop, especially for low-finesse resonators. Moreover, non-linear filtering \cite{li_broadening_2021} may be employed to reduce this noise.

\section{Experimental verification}
\label{sec:Experimental_Results}
\begin{figure}
    \includegraphics[width=0.95\linewidth]{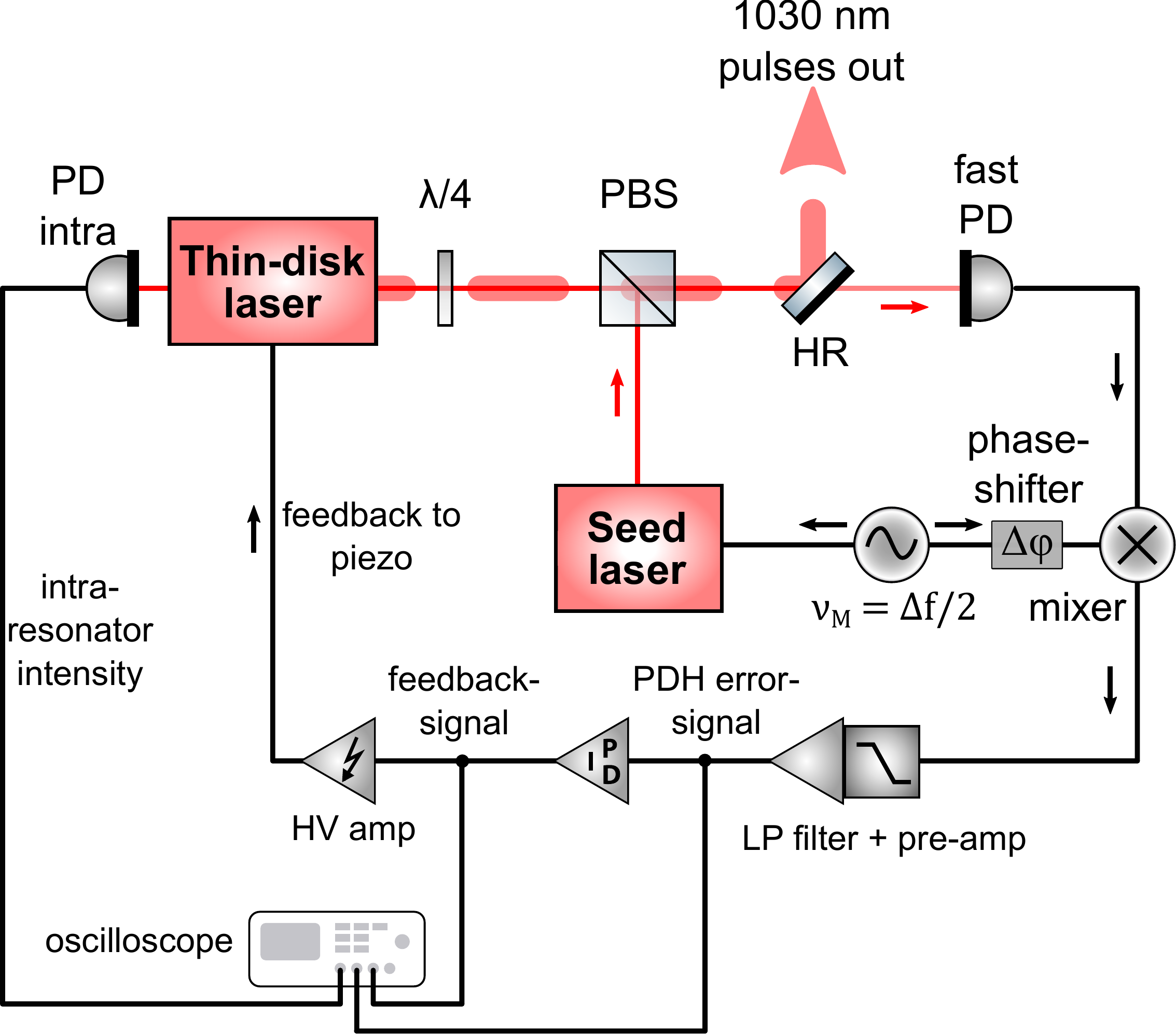}
    \caption{Sketch of the injection seeded thin-disk laser. Electronic connections are shown in black. The seed laser is current modulated at $\nu_{M}$. The PDH feedback loop, containing a mixer, a pre-amplifier and a low-pass filter (LP), outputs the error signal, which is converted into a feedback signal in the PID controller. The feedback signal is applied to a piezoelectric actuator stabilizing the length of the thin-disk laser resonator. PBS: polarizing beam splitter, HV: high voltage, $\lambda /4$: quarter-wave plate, PD: photodiode, $\Delta \varphi$: phase shifter, HR: high-reflectivity mirror.}
    \label{fig:Sketch_Osci}
\end{figure}

\begin{figure}
    \includegraphics[width=0.95\linewidth]{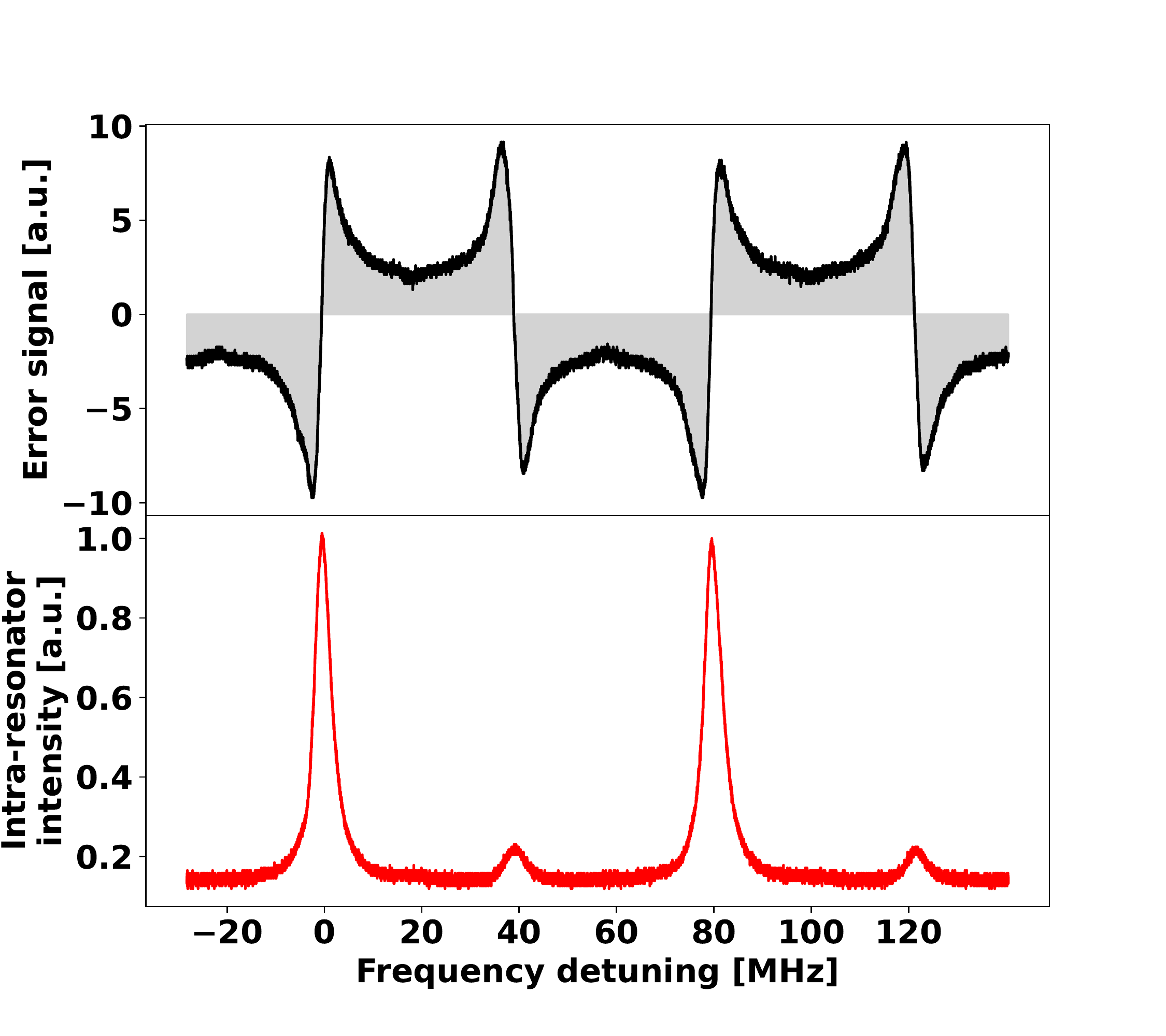}
    \caption{\textbf{Top: } Open-loop error signal of our PDH lock modulated at $\nu_{M} = \Delta f/2$ measured with a fast photodiode ("fast PD" in Fig.~\ref{fig:Sketch_Osci}). The trace was obtained by scanning the length of the seeded thin-disk laser resonator while the seed laser was modulated at $\nu_{M}=40.5$~MHz. \textbf{Bottom:} Corresponding intra-resonator intensity measured with the photodiode ("PD intra" in Fig.~\ref{fig:Sketch_Osci}) monitoring the leakage light behind an end-mirror of the resonator.}
    \label{fig:OurPDH_Measured}
\end{figure}

We tested the PDH lock at $\nu_{M}=\Delta f/2$ for an injection seeded pulsed thin-disk laser (TDL) at 1030 nm. While details on the TDL will be published elsewhere, its simplified scheme is sketched in Fig.~\ref{fig:Sketch_Osci}. A single-frequency laser with linewidth < 30 kHz (Toptica DL Pro) at 1030~nm wavelength was used to seed the TDL resonator. Resonant incoupling of the seed was achieved by stabilizing the length of the TDL resonator via the PDH method to the seed frequency. The TDL resonator length was adjusted by having one end-mirror mounted on a piezo-electric actuator (piezo). The resonator length was $L = 1.85$~m corresponding to a FSR of $\Delta f = c/2L= 81$~MHz. Phase modulating the seed laser at $\nu_{M}=\Delta f /2=40.5$~MHz thus allowed us to obtain the PDH error signal free from Trojan operating points as shown in Fig.~\ref{fig:OurPDH_Measured}. The phase modulation was achieved by current modulating the seed laser with the output from one channel of a function generator (Tektronix AFG1062). The second channel was set to the same frequency and used as a reference to demodulate the signal from the fast photodiode (PD) measuring the light reflected from the TDL resonator. The relative phase $\Delta \varphi$ between both channels could be adjusted on the function generator.

\begin{figure}
    \includegraphics[width=0.95\linewidth]{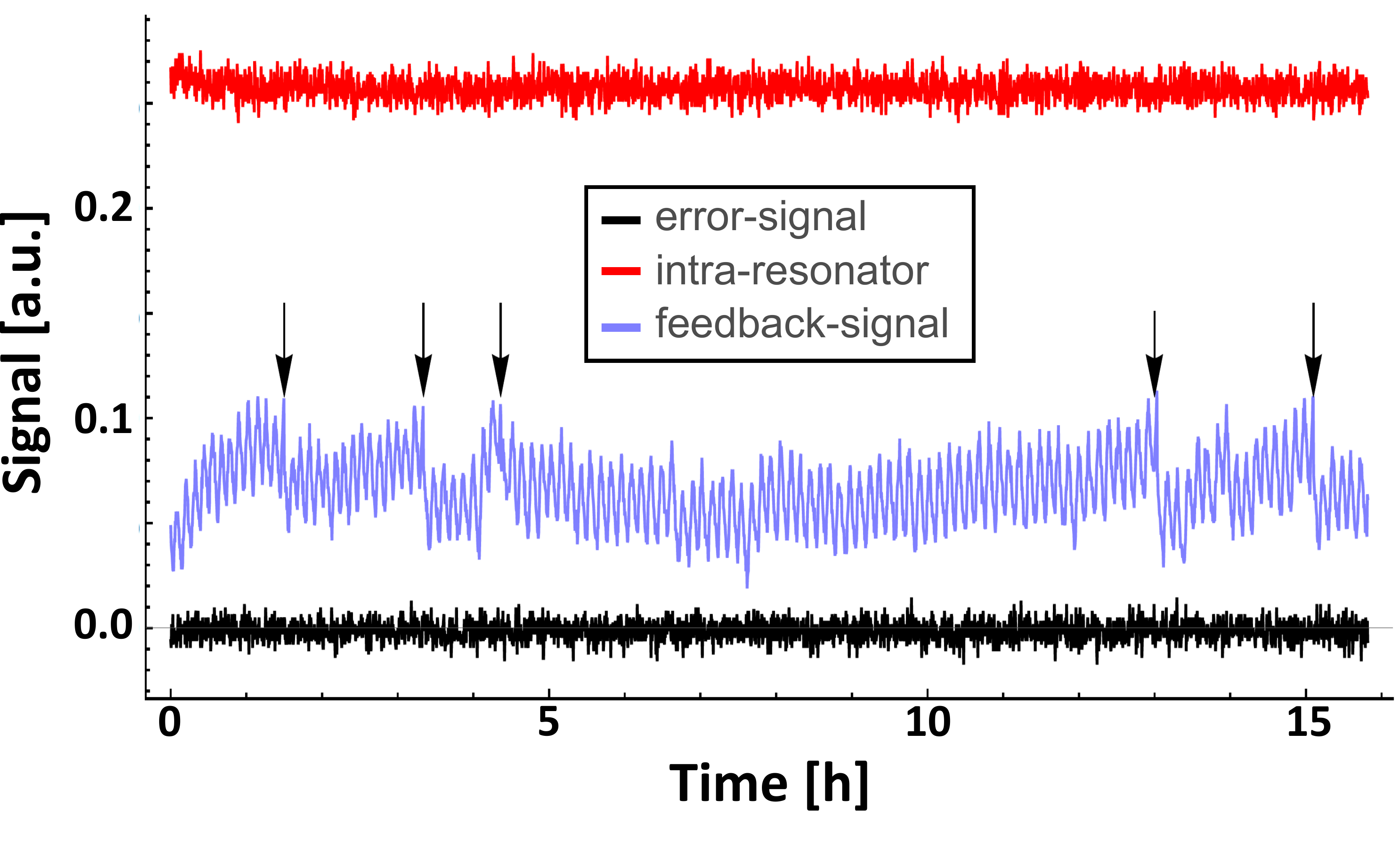}
    \caption{Measurement of the long-term stability of the PDH locked injection-seeded thin-disk laser resonator operated below lasing threshold. Closed-loop error signal, intra-resonator power and high-voltage (feedback) signal applied to the piezo element controlling the resonator length, are plotted versus time. Over the course of 15 h five relocks marked by the arrows occurred, when the PID servo-box reached its maximal output voltage. Relocking does not affect the intra-resonator power stability, demonstrating no false locking. The saw-tooth like oscillations in the feedback signal are due to temperature variations in the ambient air.}
    \label{fig:longterm_lock}
\end{figure}

We first tested the robustness of our PDH locking scheme operating the TDL below lasing threshold in CW-mode. The finesse of the TDL resonator was $\mathcal{F} \approx 19$, and its length was locked to the seed while the TDL crystal was pumped at 350~W and water impingement-cooled from the back side. Figure~\ref{fig:longterm_lock} shows the intra-resonator intensity (red), the closed loop PDH error signal (black) and the feedback voltage signal to the piezo (blue) over more than 15 hours. During this time 5 relocks occurred (marked by arrows in the figure) when the high-voltage amplifier of the PID control unit ran into its limit. No false re-locking was observed.
\begin{figure}
    \includegraphics[width=0.95\linewidth]{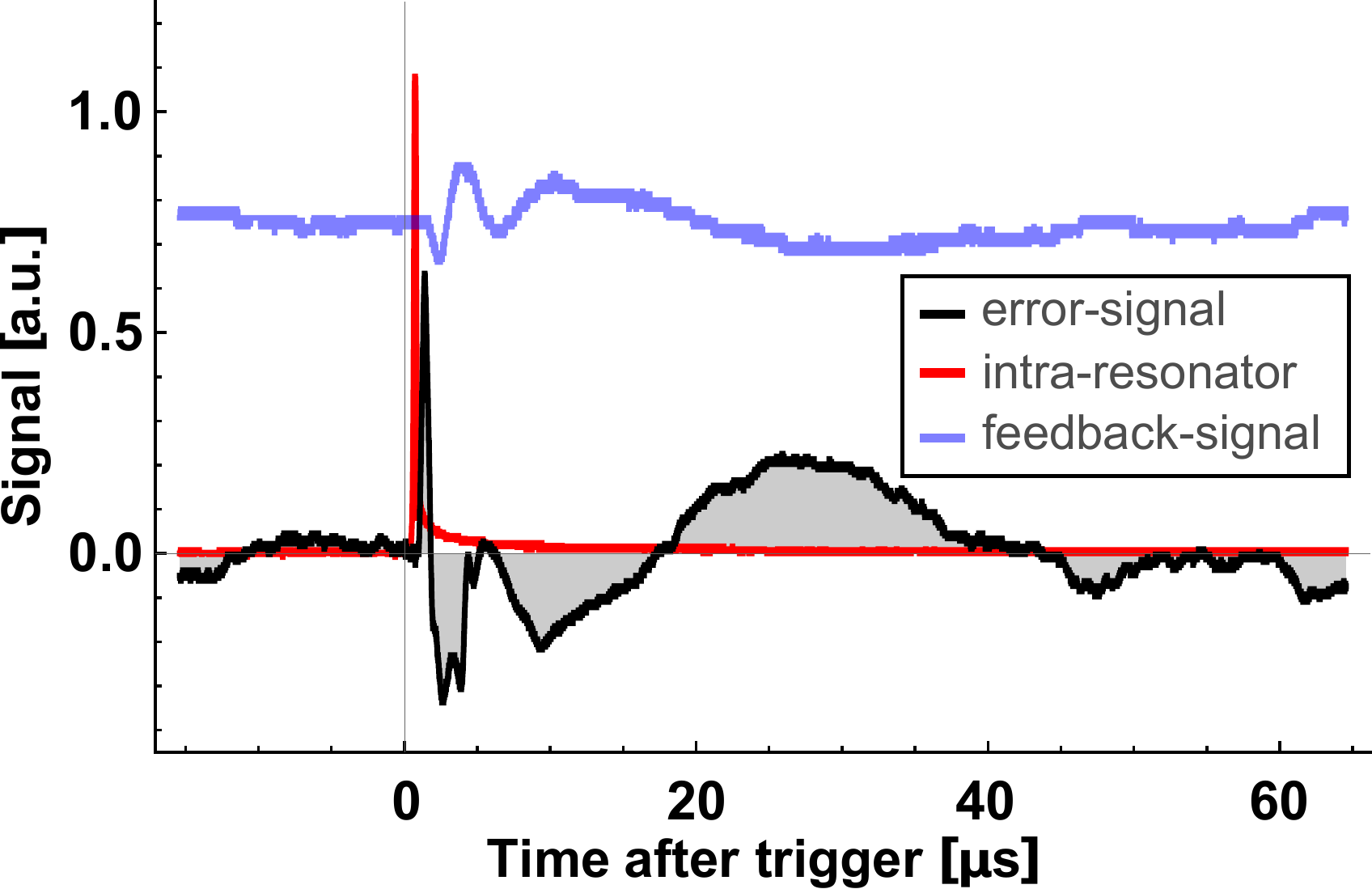}
    \caption{Measured closed-loop error signal, intra-resonator power and high-voltage (feedback) signal applied to the piezo element controlling the resonator length, versus time after pulse extraction. In this case pulses of 10~mJ energy where extracted at a pump power of 350 W. The traces are obtained from an average over 100 pulses.}
    \label{fig:relock_disk}
\end{figure}

We then tested the transient behavior of our PDH lock by seeding the TDL in CW and pulsing it with a repetition rate of 100~Hz to introduce a large disturbance in the laser dynamics. As expected, a laser pulse strongly perturbs the PDH error signal and, for a short time ($<1$~\textmu s), also saturates the fast PD. However, the saturation is beneficial since it limits the sensitivity of the feedback loop to the pulse, because the constant (saturated) signal from the PD is mixed with the reference and low-pass filtered down to zero during this time. Hence, the error signal is zero while the PD is saturated, which prevents the feedback loop from correcting the TDL resonator length by an excessive amount, essentially leaving the resonator freely drifting until the PD recovers.

Figure~\ref{fig:relock_disk} shows the intra-resonator power (red), the closed loop error signal (black), and the applied feedback voltage to the piezo (blue) during and just after a pulse extraction from our TDL. No false re-locking was observed also in this pulsed operating mode. The 20~ns long laser pulse (coupled out at $t=0$) disturbs the error signal for a short time. The fast saturation of the diode signal limits the excursion of the error signal so that the error signal stabilizes already after 40~\textmu s and the TDL resonator is again locked on resonance with the seed. Such a short stabilization time could in principle allow our system to deliver injection seeded pulses at 10-20~kHz repetition rate.


\section{Discussion and Conclusion}\label{sec:Conclusion}
We proposed a modified PDH locking scheme, where the error signal is free from Trojan operating points, by phase modulating the seed laser at half the FSR of the seeded resonator. In this way the dynamic range of the PDH lock is broadened to "infinity" (the full FSR), ensuring correct re-locking to a resonator $\textnormal{TEM}_{00}$ mode even after large perturbations.

We applied this technique to seed a thin-disk laser at 1030~nm from which we obtained single-frequency pulses of 20 ns length and 35 mJ pulse energy. The disk-laser was tested for >15~hours and no false locking was observed.

Our method requires good alignment and mode-matching of the seed to the resonator since higher-order transverse modes distort the PDH error signal and can lead to Trojan operating points. However, operating seeded laser resonators in $\textnormal{TEM}_{00}$ is not problematic thanks to aperture effects (e.g. soft aperture of the gain medium) and since typical seed lasers already run in $\textnormal{TEM}_{00}$. The reduction of the finesse due to the aperture effects is advantageous in our approach since a high finesse would lead to a PDH error signal very close to zero for detunings between the carrier and the sidebands, which, in the presence of excessive noise might trap the feedback loop and behave like a Trojan operating point.

Our locking scheme is particularly useful in situations where single-frequency operation of a master-slave system is required under rough conditions or over an extended period of time. For example, high-power/energy injection-seeded laser systems subject to large fluctuations and drifts of the running conditions would benefit from such a robust locking mechanism. Compared to widely used ramp-fire techniques \cite{lemmerz_frequency_2017,dai_development_2018,cao_theoretical_2020}, our method avoids the disadvantage of scanning the position of the end-mirror to obtain the resonance condition. Better stability and higher repetition rates might thus be feasible. Recently high-frequency RF-drivers, EOMs and photodiodes with bandwidths >~50~GHz have become available so that our technique can be applied to resonators shorter than 1~cm. This extends the range of applications of our technique to, for example, injection-seeded OPOs.

\begin{acknowledgments}
We acknowledge the support of the following grants: 
The European Research Council (ERC) through CoG. \#725039; the Swiss National Science Foundation through the projects SNF 200021\_165854 and SNF 200020\_197052; the Deutsche Forschungsgemeinschaft (DFG, German Research Foundation) under Germany's Excellence Initiative EXC 1098 PRISMA (194673446), Excellence Strategy EXC PRISMA+ (390831469) and DFG/ANR Project LASIMUS (DFG Grant Agreement 407008443); the French National Research Agency with project ANR-18-CE92-0030-02;

\end{acknowledgments}
\section*{Author Declarations}
\subsection*{Conflict of Interest}
The authors have no conflicts to disclose.
\subsection*{Author Contributions}
\noindent
\textbf{Manuel Zeyen:} Conceptualization (equal), Data curation (lead), Formal analysis (lead), Investigation (lead), Methodology (equal), Project administration (equal), Visualization (lead), Writing – original draft (lead).
\textbf{Lukas Affolter:} Data curation (equal), Investigation (supporting).
\textbf{Marwan Abdou Ahmed:} Resources (supporting).
\textbf{Thomas Graf:} Resources (supporting).
\textbf{Oguzhan Kara:} Investigation (equal).
\textbf{Klaus Kirch:} Conceptualization (supporting), Funding acquisition (equal), Project administration (supporting), Resources (equal).
\textbf{Miroslaw Marszalek:} Data curation (equal), Visualization (supporting), Writing – review \& editing (lead).
\textbf{François Nez:} Conceptualization (equal), Funding acquisition (equal).
\textbf{Ahmed Ouf:} Validation (equal), Writing – review \& editing (supporting).
\textbf{Randolf Pohl:} Funding acquisition (equal), Resources  (equal).
\textbf{Siddharth Rajamohanan:} Validation (equal).
\textbf{Pauline Yzombard:} Writing – review \ editing (supporting).
\textbf{Aldo Antognini:} Conceptualization (equal), Funding acquisition (lead), Methodology (equal), Project administration (lead), Supervision (equal).
\textbf{Karsten Schuhmann:} Conceptualization (lead), Formal analysis (supporting), Investigation (supporting), Methodology (equal), Project administration (supporting), Supervision (lead), Writing – review \& editing (supporting).

\section*{Data Availability Statement}

The data that support the findings of this study are available
from the corresponding author upon reasonable request.


\begin{thebibliography}{30}%
\makeatletter
\providecommand \@ifxundefined [1]{%
 \@ifx{#1\undefined}
}%
\providecommand \@ifnum [1]{%
 \ifnum #1\expandafter \@firstoftwo
 \else \expandafter \@secondoftwo
 \fi
}%
\providecommand \@ifx [1]{%
 \ifx #1\expandafter \@firstoftwo
 \else \expandafter \@secondoftwo
 \fi
}%
\providecommand \natexlab [1]{#1}%
\providecommand \enquote  [1]{``#1''}%
\providecommand \bibnamefont  [1]{#1}%
\providecommand \bibfnamefont [1]{#1}%
\providecommand \citenamefont [1]{#1}%
\providecommand \href@noop [0]{\@secondoftwo}%
\providecommand \href [0]{\begingroup \@sanitize@url \@href}%
\providecommand \@href[1]{\@@startlink{#1}\@@href}%
\providecommand \@@href[1]{\endgroup#1\@@endlink}%
\providecommand \@sanitize@url [0]{\catcode `\\12\catcode `\$12\catcode
  `\&12\catcode `\#12\catcode `\^12\catcode `\_12\catcode `\%12\relax}%
\providecommand \@@startlink[1]{}%
\providecommand \@@endlink[0]{}%
\providecommand \url  [0]{\begingroup\@sanitize@url \@url }%
\providecommand \@url [1]{\endgroup\@href {#1}{\urlprefix }}%
\providecommand \urlprefix  [0]{URL }%
\providecommand \Eprint [0]{\href }%
\providecommand \doibase [0]{https://doi.org/}%
\providecommand \selectlanguage [0]{\@gobble}%
\providecommand \bibinfo  [0]{\@secondoftwo}%
\providecommand \bibfield  [0]{\@secondoftwo}%
\providecommand \translation [1]{[#1]}%
\providecommand \BibitemOpen [0]{}%
\providecommand \bibitemStop [0]{}%
\providecommand \bibitemNoStop [0]{.\EOS\space}%
\providecommand \EOS [0]{\spacefactor3000\relax}%
\providecommand \BibitemShut  [1]{\csname bibitem#1\endcsname}%
\let\auto@bib@innerbib\@empty
\bibitem [{\citenamefont {Drever}\ \emph {et~al.}(1983)\citenamefont {Drever},
  \citenamefont {Hall}, \citenamefont {Kowalski}, \citenamefont {Hough},
  \citenamefont {Ford}, \citenamefont {Munley},\ and\ \citenamefont
  {Ward}}]{drever_laser_1983}%
  \BibitemOpen
  \bibfield  {author} {\bibinfo {author} {\bibfnamefont {R.~W.~P.}\
  \bibnamefont {Drever}}, \bibinfo {author} {\bibfnamefont {J.~L.}\
  \bibnamefont {Hall}}, \bibinfo {author} {\bibfnamefont {F.~V.}\ \bibnamefont
  {Kowalski}}, \bibinfo {author} {\bibfnamefont {J.}~\bibnamefont {Hough}},
  \bibinfo {author} {\bibfnamefont {G.~M.}\ \bibnamefont {Ford}}, \bibinfo
  {author} {\bibfnamefont {A.~J.}\ \bibnamefont {Munley}},\ and\ \bibinfo
  {author} {\bibfnamefont {H.}~\bibnamefont {Ward}},\ }\bibfield  {title}
  {\enquote {\bibinfo {title} {Laser phase and frequency stabilization using an
  optical resonator},}\ }\href {https://doi.org/10.1007/BF00702605} {\bibfield
  {journal} {\bibinfo  {journal} {Applied Physics B Photophysics and Laser
  Chemistry}\ }\textbf {\bibinfo {volume} {31}},\ \bibinfo {pages} {97--105}
  (\bibinfo {year} {1983})}\BibitemShut {NoStop}%
\bibitem [{\citenamefont {Kessler}\ \emph {et~al.}(2012)\citenamefont
  {Kessler}, \citenamefont {Hagemann}, \citenamefont {Grebing}, \citenamefont
  {Legero}, \citenamefont {Sterr}, \citenamefont {Riehle}, \citenamefont
  {Martin}, \citenamefont {Chen},\ and\ \citenamefont
  {Ye}}]{kessler_sub-40-mhz-linewidth_2012}%
  \BibitemOpen
  \bibfield  {author} {\bibinfo {author} {\bibfnamefont {T.}~\bibnamefont
  {Kessler}}, \bibinfo {author} {\bibfnamefont {C.}~\bibnamefont {Hagemann}},
  \bibinfo {author} {\bibfnamefont {C.}~\bibnamefont {Grebing}}, \bibinfo
  {author} {\bibfnamefont {T.}~\bibnamefont {Legero}}, \bibinfo {author}
  {\bibfnamefont {U.}~\bibnamefont {Sterr}}, \bibinfo {author} {\bibfnamefont
  {F.}~\bibnamefont {Riehle}}, \bibinfo {author} {\bibfnamefont {M.~J.}\
  \bibnamefont {Martin}}, \bibinfo {author} {\bibfnamefont {L.}~\bibnamefont
  {Chen}},\ and\ \bibinfo {author} {\bibfnamefont {J.}~\bibnamefont {Ye}},\
  }\bibfield  {title} {\enquote {\bibinfo {title} {A sub-40-{mHz}-linewidth
  laser based on a silicon single-crystal optical cavity},}\ }\href
  {https://doi.org/10.1038/nphoton.2012.217} {\bibfield  {journal} {\bibinfo
  {journal} {Nature Photonics}\ }\textbf {\bibinfo {volume} {6}},\ \bibinfo
  {pages} {687--692} (\bibinfo {year} {2012})}\BibitemShut {NoStop}%
\bibitem [{\citenamefont {Schmid}\ \emph {et~al.}(2019)\citenamefont {Schmid},
  \citenamefont {Weitenberg}, \citenamefont {Hänsch}, \citenamefont {Udem},\
  and\ \citenamefont {Ozawa}}]{schmid_simple_2019}%
  \BibitemOpen
  \bibfield  {author} {\bibinfo {author} {\bibfnamefont {F.}~\bibnamefont
  {Schmid}}, \bibinfo {author} {\bibfnamefont {J.}~\bibnamefont {Weitenberg}},
  \bibinfo {author} {\bibfnamefont {T.~W.}\ \bibnamefont {Hänsch}}, \bibinfo
  {author} {\bibfnamefont {T.}~\bibnamefont {Udem}},\ and\ \bibinfo {author}
  {\bibfnamefont {A.}~\bibnamefont {Ozawa}},\ }\bibfield  {title} {\enquote
  {\bibinfo {title} {Simple phase noise measurement scheme for
  cavity-stabilized laser systems},}\ }\href
  {https://doi.org/10.1364/OL.44.002709} {\bibfield  {journal} {\bibinfo
  {journal} {Optics Letters}\ }\textbf {\bibinfo {volume} {44}},\ \bibinfo
  {pages} {2709--2712} (\bibinfo {year} {2019})}\BibitemShut {NoStop}%
\bibitem [{\citenamefont {Willke}(2010)}]{willke_stabilized_2010}%
  \BibitemOpen
  \bibfield  {author} {\bibinfo {author} {\bibfnamefont {B.}~\bibnamefont
  {Willke}},\ }\bibfield  {title} {\enquote {\bibinfo {title} {Stabilized
  lasers for advanced gravitational wave detectors},}\ }\href
  {https://doi.org/10.1002/lpor.200900036} {\bibfield  {journal} {\bibinfo
  {journal} {Laser \& Photonics Reviews}\ }\textbf {\bibinfo {volume} {4}},\
  \bibinfo {pages} {780--794} (\bibinfo {year} {2010})}\BibitemShut {NoStop}%
\bibitem [{\citenamefont {Hond}\ \emph {et~al.}(2017)\citenamefont {Hond},
  \citenamefont {Cisternas}, \citenamefont {Lochead},\ and\ \citenamefont
  {Druten}}]{hond_medium-finesse_2017}%
  \BibitemOpen
  \bibfield  {author} {\bibinfo {author} {\bibfnamefont {J.~d.}\ \bibnamefont
  {Hond}}, \bibinfo {author} {\bibfnamefont {N.}~\bibnamefont {Cisternas}},
  \bibinfo {author} {\bibfnamefont {G.}~\bibnamefont {Lochead}},\ and\ \bibinfo
  {author} {\bibfnamefont {N.~J.~v.}\ \bibnamefont {Druten}},\ }\bibfield
  {title} {\enquote {\bibinfo {title} {Medium-finesse optical cavity for the
  stabilization of {Rydberg} lasers},}\ }\href
  {https://doi.org/10.1364/AO.56.005436} {\bibfield  {journal} {\bibinfo
  {journal} {Applied Optics}\ }\textbf {\bibinfo {volume} {56}},\ \bibinfo
  {pages} {5436--5443} (\bibinfo {year} {2017})}\BibitemShut {NoStop}%
\bibitem [{\citenamefont {Legaie}, \citenamefont {Picken},\ and\ \citenamefont
  {Pritchard}(2018)}]{legaie_sub-kilohertz_2018}%
  \BibitemOpen
  \bibfield  {author} {\bibinfo {author} {\bibfnamefont {R.}~\bibnamefont
  {Legaie}}, \bibinfo {author} {\bibfnamefont {C.~J.}\ \bibnamefont {Picken}},\
  and\ \bibinfo {author} {\bibfnamefont {J.~D.}\ \bibnamefont {Pritchard}},\
  }\bibfield  {title} {\enquote {\bibinfo {title} {Sub-kilohertz excitation
  lasers for quantum information processing with {Rydberg} atoms},}\ }\href
  {https://doi.org/10.1364/JOSAB.35.000892} {\bibfield  {journal} {\bibinfo
  {journal} {JOSA B}\ }\textbf {\bibinfo {volume} {35}},\ \bibinfo {pages}
  {892--898} (\bibinfo {year} {2018})}\BibitemShut {NoStop}%
\bibitem [{\citenamefont {Grinin}\ \emph {et~al.}(2020)\citenamefont {Grinin},
  \citenamefont {Matveev}, \citenamefont {Yost}, \citenamefont {Maisenbacher},
  \citenamefont {Wirthl}, \citenamefont {Pohl}, \citenamefont {Hänsch},\ and\
  \citenamefont {Udem}}]{grinin_two-photon_2020}%
  \BibitemOpen
  \bibfield  {author} {\bibinfo {author} {\bibfnamefont {A.}~\bibnamefont
  {Grinin}}, \bibinfo {author} {\bibfnamefont {A.}~\bibnamefont {Matveev}},
  \bibinfo {author} {\bibfnamefont {D.~C.}\ \bibnamefont {Yost}}, \bibinfo
  {author} {\bibfnamefont {L.}~\bibnamefont {Maisenbacher}}, \bibinfo {author}
  {\bibfnamefont {V.}~\bibnamefont {Wirthl}}, \bibinfo {author} {\bibfnamefont
  {R.}~\bibnamefont {Pohl}}, \bibinfo {author} {\bibfnamefont {T.~W.}\
  \bibnamefont {Hänsch}},\ and\ \bibinfo {author} {\bibfnamefont
  {T.}~\bibnamefont {Udem}},\ }\bibfield  {title} {\enquote {\bibinfo {title}
  {Two-photon frequency comb spectroscopy of atomic hydrogen},}\ }\href
  {https://doi.org/10.1126/science.abc7776} {\bibfield  {journal} {\bibinfo
  {journal} {Science}\ }\textbf {\bibinfo {volume} {370}},\ \bibinfo {pages}
  {1061--1066} (\bibinfo {year} {2020})}\BibitemShut {NoStop}%
\bibitem [{\citenamefont {Wulfmeyer}\ \emph {et~al.}(2000)\citenamefont
  {Wulfmeyer}, \citenamefont {Randall}, \citenamefont {Brewer},\ and\
  \citenamefont {Hardesty}}]{wulfmeyer_2-m_2000}%
  \BibitemOpen
  \bibfield  {author} {\bibinfo {author} {\bibfnamefont {V.}~\bibnamefont
  {Wulfmeyer}}, \bibinfo {author} {\bibfnamefont {M.}~\bibnamefont {Randall}},
  \bibinfo {author} {\bibfnamefont {A.}~\bibnamefont {Brewer}},\ and\ \bibinfo
  {author} {\bibfnamefont {R.~M.}\ \bibnamefont {Hardesty}},\ }\bibfield
  {title} {\enquote {\bibinfo {title} {2-µm {Doppler} lidar transmitter with
  high frequency stability and low chirp},}\ }\href
  {https://doi.org/10.1364/OL.25.001228} {\bibfield  {journal} {\bibinfo
  {journal} {Optics Letters}\ }\textbf {\bibinfo {volume} {25}},\ \bibinfo
  {pages} {1228--1230} (\bibinfo {year} {2000})}\BibitemShut {NoStop}%
\bibitem [{\citenamefont {Chen}, \citenamefont {Gao},\ and\ \citenamefont
  {Wang}(2022)}]{chen_injection-seeded_2022}%
  \BibitemOpen
  \bibfield  {author} {\bibinfo {author} {\bibfnamefont {C.}~\bibnamefont
  {Chen}}, \bibinfo {author} {\bibfnamefont {C.}~\bibnamefont {Gao}},\ and\
  \bibinfo {author} {\bibfnamefont {Q.}~\bibnamefont {Wang}},\ }\bibfield
  {title} {\enquote {\bibinfo {title} {Injection-seeded 10 {kHz} repetition
  rate {Er}:{YAG} solid-state laser with single-frequency pulse energy more
  than 1 {mJ}},}\ }\href {https://doi.org/10.1364/OE.458583} {\bibfield
  {journal} {\bibinfo  {journal} {Optics Express}\ }\textbf {\bibinfo {volume}
  {30}},\ \bibinfo {pages} {16044--16052} (\bibinfo {year} {2022})}\BibitemShut
  {NoStop}%
\bibitem [{\citenamefont {Zhang}\ \emph {et~al.}(2020)\citenamefont {Zhang},
  \citenamefont {Tan}, \citenamefont {Wang}, \citenamefont {Cheng},
  \citenamefont {Sun}, \citenamefont {Liu},\ and\ \citenamefont
  {Hu}}]{zhang_seeded_2020}%
  \BibitemOpen
  \bibfield  {author} {\bibinfo {author} {\bibfnamefont {Z.-T.}\ \bibnamefont
  {Zhang}}, \bibinfo {author} {\bibfnamefont {Y.}~\bibnamefont {Tan}}, \bibinfo
  {author} {\bibfnamefont {J.}~\bibnamefont {Wang}}, \bibinfo {author}
  {\bibfnamefont {C.-F.}\ \bibnamefont {Cheng}}, \bibinfo {author}
  {\bibfnamefont {Y.~R.}\ \bibnamefont {Sun}}, \bibinfo {author} {\bibfnamefont
  {A.-W.}\ \bibnamefont {Liu}},\ and\ \bibinfo {author} {\bibfnamefont {S.-M.}\
  \bibnamefont {Hu}},\ }\bibfield  {title} {\enquote {\bibinfo {title} {Seeded
  optical parametric oscillator light source for precision spectroscopy},}\
  }\href {https://doi.org/10.1364/OL.384582} {\bibfield  {journal} {\bibinfo
  {journal} {Optics Letters}\ }\textbf {\bibinfo {volume} {45}},\ \bibinfo
  {pages} {1013} (\bibinfo {year} {2020})}\BibitemShut {NoStop}%
\bibitem [{\citenamefont {Ricciardi}\ \emph {et~al.}(2015)\citenamefont
  {Ricciardi}, \citenamefont {Mosca}, \citenamefont {Parisi}, \citenamefont
  {Maddaloni}, \citenamefont {Santamaria}, \citenamefont {Natale},\ and\
  \citenamefont {Rosa}}]{ricciardi_sub-kilohertz_2015}%
  \BibitemOpen
  \bibfield  {author} {\bibinfo {author} {\bibfnamefont {I.}~\bibnamefont
  {Ricciardi}}, \bibinfo {author} {\bibfnamefont {S.}~\bibnamefont {Mosca}},
  \bibinfo {author} {\bibfnamefont {M.}~\bibnamefont {Parisi}}, \bibinfo
  {author} {\bibfnamefont {P.}~\bibnamefont {Maddaloni}}, \bibinfo {author}
  {\bibfnamefont {L.}~\bibnamefont {Santamaria}}, \bibinfo {author}
  {\bibfnamefont {P.~D.}\ \bibnamefont {Natale}},\ and\ \bibinfo {author}
  {\bibfnamefont {M.~D.}\ \bibnamefont {Rosa}},\ }\bibfield  {title} {\enquote
  {\bibinfo {title} {Sub-kilohertz linewidth narrowing of a mid-infrared
  optical parametric oscillator idler frequency by direct cavity
  stabilization},}\ }\href {https://doi.org/10.1364/OL.40.004743} {\bibfield
  {journal} {\bibinfo  {journal} {Optics Letters}\ }\textbf {\bibinfo {volume}
  {40}},\ \bibinfo {pages} {4743--4746} (\bibinfo {year} {2015})}\BibitemShut
  {NoStop}%
\bibitem [{\citenamefont {Cygan}\ \emph {et~al.}(2011)\citenamefont {Cygan},
  \citenamefont {Lisak}, \citenamefont {Wójtewicz}, \citenamefont
  {Domysławska}, \citenamefont {Trawiński},\ and\ \citenamefont
  {Ciuryło}}]{cygan_active_2011}%
  \BibitemOpen
  \bibfield  {author} {\bibinfo {author} {\bibfnamefont {A.}~\bibnamefont
  {Cygan}}, \bibinfo {author} {\bibfnamefont {D.}~\bibnamefont {Lisak}},
  \bibinfo {author} {\bibfnamefont {S.}~\bibnamefont {Wójtewicz}}, \bibinfo
  {author} {\bibfnamefont {J.}~\bibnamefont {Domysławska}}, \bibinfo {author}
  {\bibfnamefont {R.~S.}\ \bibnamefont {Trawiński}},\ and\ \bibinfo {author}
  {\bibfnamefont {R.}~\bibnamefont {Ciuryło}},\ }\bibfield  {title} {\enquote
  {\bibinfo {title} {Active control of the {Pound}–{Drever}–{Hall} error
  signal offset in high-repetition-rate cavity ring-down spectroscopy},}\
  }\href {https://doi.org/10.1088/0957-0233/22/11/115303} {\bibfield  {journal}
  {\bibinfo  {journal} {Measurement Science and Technology}\ }\textbf {\bibinfo
  {volume} {22}},\ \bibinfo {pages} {115303} (\bibinfo {year}
  {2011})}\BibitemShut {NoStop}%
\bibitem [{\citenamefont {Gatti}\ \emph {et~al.}(2015)\citenamefont {Gatti},
  \citenamefont {Gotti}, \citenamefont {Sala}, \citenamefont {Coluccelli},
  \citenamefont {Belmonte}, \citenamefont {Prevedelli}, \citenamefont
  {Laporta},\ and\ \citenamefont {Marangoni}}]{gatti_wide-bandwidth_2015}%
  \BibitemOpen
  \bibfield  {author} {\bibinfo {author} {\bibfnamefont {D.}~\bibnamefont
  {Gatti}}, \bibinfo {author} {\bibfnamefont {R.}~\bibnamefont {Gotti}},
  \bibinfo {author} {\bibfnamefont {T.}~\bibnamefont {Sala}}, \bibinfo {author}
  {\bibfnamefont {N.}~\bibnamefont {Coluccelli}}, \bibinfo {author}
  {\bibfnamefont {M.}~\bibnamefont {Belmonte}}, \bibinfo {author}
  {\bibfnamefont {M.}~\bibnamefont {Prevedelli}}, \bibinfo {author}
  {\bibfnamefont {P.}~\bibnamefont {Laporta}},\ and\ \bibinfo {author}
  {\bibfnamefont {M.}~\bibnamefont {Marangoni}},\ }\bibfield  {title} {\enquote
  {\bibinfo {title} {Wide-bandwidth {Pound}–{Drever}–{Hall} locking through
  a single-sideband modulator},}\ }\href {https://doi.org/10.1364/OL.40.005176}
  {\bibfield  {journal} {\bibinfo  {journal} {Optics Letters}\ }\textbf
  {\bibinfo {volume} {40}},\ \bibinfo {pages} {5176} (\bibinfo {year}
  {2015})}\BibitemShut {NoStop}%
\bibitem [{\citenamefont {Izumi}, \citenamefont {Sigg},\ and\ \citenamefont
  {Barsotti}()}]{izumi_self_nodate}%
  \BibitemOpen
  \bibfield  {author} {\bibinfo {author} {\bibfnamefont {K.}~\bibnamefont
  {Izumi}}, \bibinfo {author} {\bibfnamefont {D.}~\bibnamefont {Sigg}},\ and\
  \bibinfo {author} {\bibfnamefont {L.}~\bibnamefont {Barsotti}},\ }\bibfield
  {title} {\enquote {\bibinfo {title} {Self {Ampliﬁed} {Lock} of a
  {Ultra}-narrow {Linewidth} {Optical} {Cavity}},}\ }\href@noop {} {\ ,\
  \bibinfo {pages} {5}}\BibitemShut {NoStop}%
\bibitem [{\citenamefont {Zeng}\ \emph {et~al.}(2021)\citenamefont {Zeng},
  \citenamefont {Fu}, \citenamefont {Liu}, \citenamefont {He}, \citenamefont
  {Liu}, \citenamefont {Xu}, \citenamefont {Sun},\ and\ \citenamefont
  {Wang}}]{zeng_stabilizing_2021}%
  \BibitemOpen
  \bibfield  {author} {\bibinfo {author} {\bibfnamefont {Y.}~\bibnamefont
  {Zeng}}, \bibinfo {author} {\bibfnamefont {Z.}~\bibnamefont {Fu}}, \bibinfo
  {author} {\bibfnamefont {Y.-Y.}\ \bibnamefont {Liu}}, \bibinfo {author}
  {\bibfnamefont {X.-D.}\ \bibnamefont {He}}, \bibinfo {author} {\bibfnamefont
  {M.}~\bibnamefont {Liu}}, \bibinfo {author} {\bibfnamefont {P.}~\bibnamefont
  {Xu}}, \bibinfo {author} {\bibfnamefont {X.-H.}\ \bibnamefont {Sun}},\ and\
  \bibinfo {author} {\bibfnamefont {J.}~\bibnamefont {Wang}},\ }\bibfield
  {title} {\enquote {\bibinfo {title} {Stabilizing a laser frequency by the
  {Pound}–{Drever}–{Hall} technique with an acousto-optic modulator},}\
  }\href {https://doi.org/10.1364/AO.415011} {\bibfield  {journal} {\bibinfo
  {journal} {Applied Optics}\ }\textbf {\bibinfo {volume} {60}},\ \bibinfo
  {pages} {1159} (\bibinfo {year} {2021})}\BibitemShut {NoStop}%
\bibitem [{\citenamefont {Wang}\ \emph {et~al.}(2019)\citenamefont {Wang},
  \citenamefont {Xu}, \citenamefont {Ma}, \citenamefont {Cai}, \citenamefont
  {You},\ and\ \citenamefont {Liu}}]{wang_artificial_2019}%
  \BibitemOpen
  \bibfield  {author} {\bibinfo {author} {\bibfnamefont {H.~M.}\ \bibnamefont
  {Wang}}, \bibinfo {author} {\bibfnamefont {Z.~S.}\ \bibnamefont {Xu}},
  \bibinfo {author} {\bibfnamefont {S.~C.}\ \bibnamefont {Ma}}, \bibinfo
  {author} {\bibfnamefont {M.~H.}\ \bibnamefont {Cai}}, \bibinfo {author}
  {\bibfnamefont {S.~H.}\ \bibnamefont {You}},\ and\ \bibinfo {author}
  {\bibfnamefont {H.~P.}\ \bibnamefont {Liu}},\ }\bibfield  {title} {\enquote
  {\bibinfo {title} {Artificial modulation-free {Pound}–{Drever}–{Hall}
  method for laser frequency stabilization},}\ }\href
  {https://doi.org/10.1364/OL.44.005816} {\bibfield  {journal} {\bibinfo
  {journal} {Optics Letters}\ }\textbf {\bibinfo {volume} {44}},\ \bibinfo
  {pages} {5816} (\bibinfo {year} {2019})}\BibitemShut {NoStop}%
\bibitem [{\citenamefont {Homer}()}]{homer_odyssey_nodate}%
  \BibitemOpen
  \bibfield  {author} {\bibinfo {author} {\bibnamefont {Homer}},\ }\href@noop
  {} {\emph {\bibinfo {title} {The {Odyssey}}}},\ \bibinfo {note} {scroll 4,
  line 21}\BibitemShut {NoStop}%
\bibitem [{\citenamefont {{Yen-Ting Wang}}, \citenamefont {{Degang Chen}},\
  and\ \citenamefont {Geiger}(2013)}]{yen-ting_wang_practical_2013}%
  \BibitemOpen
  \bibfield  {author} {\bibinfo {author} {\bibnamefont {{Yen-Ting Wang}}},
  \bibinfo {author} {\bibnamefont {{Degang Chen}}},\ and\ \bibinfo {author}
  {\bibfnamefont {R.~L.}\ \bibnamefont {Geiger}},\ }\bibfield  {title}
  {\enquote {\bibinfo {title} {Practical methods for verifying removal of
  {Trojan} stable operating points},}\ }in\ \href
  {https://doi.org/10.1109/ISCAS.2013.6572425} {\emph {\bibinfo {booktitle}
  {2013 {IEEE} {International} {Symposium} on {Circuits} and {Systems}
  ({ISCAS2013})}}}\ (\bibinfo  {publisher} {IEEE},\ \bibinfo {address}
  {Beijing},\ \bibinfo {year} {2013})\ pp.\ \bibinfo {pages}
  {2658--2661}\BibitemShut {NoStop}%
\bibitem [{\citenamefont {Schütte}\ \emph {et~al.}(2016)\citenamefont
  {Schütte}, \citenamefont {Hassen}, \citenamefont {Karvinen}, \citenamefont
  {Boyson}, \citenamefont {Kallapur}, \citenamefont {Song}, \citenamefont
  {Petersen}, \citenamefont {Huntington},\ and\ \citenamefont
  {Heurs}}]{schutte_experimental_2016}%
  \BibitemOpen
  \bibfield  {author} {\bibinfo {author} {\bibfnamefont {D.}~\bibnamefont
  {Schütte}}, \bibinfo {author} {\bibfnamefont {S.~S.}\ \bibnamefont
  {Hassen}}, \bibinfo {author} {\bibfnamefont {K.~S.}\ \bibnamefont
  {Karvinen}}, \bibinfo {author} {\bibfnamefont {T.~K.}\ \bibnamefont
  {Boyson}}, \bibinfo {author} {\bibfnamefont {A.~G.}\ \bibnamefont
  {Kallapur}}, \bibinfo {author} {\bibfnamefont {H.}~\bibnamefont {Song}},
  \bibinfo {author} {\bibfnamefont {I.~R.}\ \bibnamefont {Petersen}}, \bibinfo
  {author} {\bibfnamefont {E.~H.}\ \bibnamefont {Huntington}},\ and\ \bibinfo
  {author} {\bibfnamefont {M.}~\bibnamefont {Heurs}},\ }\bibfield  {title}
  {\enquote {\bibinfo {title} {Experimental {Demonstration} of {Frequency}
  {Autolocking} an {Optical} {Cavity} {Using} a {Time}-{Varying} {Kalman}
  {Filter}},}\ }\href {https://doi.org/10.1103/PhysRevApplied.5.014005}
  {\bibfield  {journal} {\bibinfo  {journal} {Physical Review Applied}\
  }\textbf {\bibinfo {volume} {5}},\ \bibinfo {pages} {014005} (\bibinfo {year}
  {2016})}\BibitemShut {NoStop}%
\bibitem [{\citenamefont {Haze}\ \emph {et~al.}(2013)\citenamefont {Haze},
  \citenamefont {Hata}, \citenamefont {Fujinaga},\ and\ \citenamefont
  {Mukaiyama}}]{haze_note_2013}%
  \BibitemOpen
  \bibfield  {author} {\bibinfo {author} {\bibfnamefont {S.}~\bibnamefont
  {Haze}}, \bibinfo {author} {\bibfnamefont {S.}~\bibnamefont {Hata}}, \bibinfo
  {author} {\bibfnamefont {M.}~\bibnamefont {Fujinaga}},\ and\ \bibinfo
  {author} {\bibfnamefont {T.}~\bibnamefont {Mukaiyama}},\ }\bibfield  {title}
  {\enquote {\bibinfo {title} {Note: {Auto}-relock system for a bow-tie cavity
  for second harmonic generation},}\ }\href {https://doi.org/10.1063/1.4793613}
  {\bibfield  {journal} {\bibinfo  {journal} {Review of Scientific
  Instruments}\ }\textbf {\bibinfo {volume} {84}},\ \bibinfo {pages} {026111}
  (\bibinfo {year} {2013})}\BibitemShut {NoStop}%
\bibitem [{\citenamefont {Guo}\ \emph {et~al.}(2022)\citenamefont {Guo},
  \citenamefont {Zhang}, \citenamefont {Liu}, \citenamefont {Chen},
  \citenamefont {Fan}, \citenamefont {Xu}, \citenamefont {Liu}, \citenamefont
  {Dong},\ and\ \citenamefont {Zhang}}]{guo_automatic_2022}%
  \BibitemOpen
  \bibfield  {author} {\bibinfo {author} {\bibfnamefont {X.}~\bibnamefont
  {Guo}}, \bibinfo {author} {\bibfnamefont {L.}~\bibnamefont {Zhang}}, \bibinfo
  {author} {\bibfnamefont {J.}~\bibnamefont {Liu}}, \bibinfo {author}
  {\bibfnamefont {L.}~\bibnamefont {Chen}}, \bibinfo {author} {\bibfnamefont
  {L.}~\bibnamefont {Fan}}, \bibinfo {author} {\bibfnamefont {G.}~\bibnamefont
  {Xu}}, \bibinfo {author} {\bibfnamefont {T.}~\bibnamefont {Liu}}, \bibinfo
  {author} {\bibfnamefont {R.}~\bibnamefont {Dong}},\ and\ \bibinfo {author}
  {\bibfnamefont {S.}~\bibnamefont {Zhang}},\ }\bibfield  {title} {\enquote
  {\bibinfo {title} {An automatic frequency stabilized laser with hertz-level
  linewidth},}\ }\href {https://doi.org/10.1016/j.optlastec.2021.107498}
  {\bibfield  {journal} {\bibinfo  {journal} {Optics \& Laser Technology}\
  }\textbf {\bibinfo {volume} {145}},\ \bibinfo {pages} {107498} (\bibinfo
  {year} {2022})}\BibitemShut {NoStop}%
\bibitem [{\citenamefont {Evans}(2002)}]{evans_lock_2002}%
  \BibitemOpen
  \bibfield  {author} {\bibinfo {author} {\bibfnamefont {M.~J.}\ \bibnamefont
  {Evans}},\ }\emph {\bibinfo {title} {Lock {Acquisition} in {Resonant}
  {Optical} {Interferometers}}},\ \href {https://doi.org/10.7907/N1J2-M098}
  {\bibinfo {type} {phd}},\ \bibinfo  {school} {California Institute of
  Technology} (\bibinfo {year} {2002})\BibitemShut {NoStop}%
\bibitem [{\citenamefont {Li}\ \emph {et~al.}(2021)\citenamefont {Li},
  \citenamefont {Wang}, \citenamefont {Dmitriev}, \citenamefont {Maggiore},
  \citenamefont {Miao},\ and\ \citenamefont {Han}}]{li_broadening_2021}%
  \BibitemOpen
  \bibfield  {author} {\bibinfo {author} {\bibfnamefont {C.}~\bibnamefont
  {Li}}, \bibinfo {author} {\bibfnamefont {H.}~\bibnamefont {Wang}}, \bibinfo
  {author} {\bibfnamefont {A.}~\bibnamefont {Dmitriev}}, \bibinfo {author}
  {\bibfnamefont {R.}~\bibnamefont {Maggiore}}, \bibinfo {author}
  {\bibfnamefont {H.}~\bibnamefont {Miao}},\ and\ \bibinfo {author}
  {\bibfnamefont {S.}~\bibnamefont {Han}},\ }\bibfield  {title} {\enquote
  {\bibinfo {title} {Broadening the dynamic range of the
  {Pound}–{Drever}–{Hall} frequency stabilization technique},}\ }\href
  {https://doi.org/10.1016/j.rinp.2021.104835} {\bibfield  {journal} {\bibinfo
  {journal} {Results in Physics}\ }\textbf {\bibinfo {volume} {30}},\ \bibinfo
  {pages} {104835} (\bibinfo {year} {2021})}\BibitemShut {NoStop}%
\bibitem [{\citenamefont {Miyoki}, \citenamefont {Telada},\ and\ \citenamefont
  {Uchiyama}(2010)}]{miyoki_expansion_2010}%
  \BibitemOpen
  \bibfield  {author} {\bibinfo {author} {\bibfnamefont {S.}~\bibnamefont
  {Miyoki}}, \bibinfo {author} {\bibfnamefont {S.}~\bibnamefont {Telada}},\
  and\ \bibinfo {author} {\bibfnamefont {T.}~\bibnamefont {Uchiyama}},\
  }\bibfield  {title} {\enquote {\bibinfo {title} {Expansion of linear range of
  {Pound}-{Drever}-{Hall} signal},}\ }\href
  {https://doi.org/10.1364/AO.49.005217} {\bibfield  {journal} {\bibinfo
  {journal} {Applied Optics}\ }\textbf {\bibinfo {volume} {49}},\ \bibinfo
  {pages} {5217--5225} (\bibinfo {year} {2010})}\BibitemShut {NoStop}%
\bibitem [{\citenamefont {Pollnau}\ and\ \citenamefont
  {Eichhorn}(2020)}]{pollnau_spectral_2020}%
  \BibitemOpen
  \bibfield  {author} {\bibinfo {author} {\bibfnamefont {M.}~\bibnamefont
  {Pollnau}}\ and\ \bibinfo {author} {\bibfnamefont {M.}~\bibnamefont
  {Eichhorn}},\ }\bibfield  {title} {\enquote {\bibinfo {title} {Spectral
  coherence, {Part} {I}: {Passive}-resonator linewidth, fundamental laser
  linewidth, and {Schawlow}-{Townes} approximation},}\ }\href
  {https://doi.org/10.1016/j.pquantelec.2020.100255} {\bibfield  {journal}
  {\bibinfo  {journal} {Progress in Quantum Electronics}\ }\textbf {\bibinfo
  {volume} {72}},\ \bibinfo {pages} {100255} (\bibinfo {year}
  {2020})}\BibitemShut {NoStop}%
\bibitem [{\citenamefont {Siegman}(1986)}]{siegman_lasers_1986}%
  \BibitemOpen
  \bibfield  {author} {\bibinfo {author} {\bibfnamefont {A.}~\bibnamefont
  {Siegman}},\ }\href@noop {} {\emph {\bibinfo {title} {Lasers}}}\ (\bibinfo
  {publisher} {University Science Books},\ \bibinfo {address} {Sausalito,
  California},\ \bibinfo {year} {1986})\BibitemShut {NoStop}%
\bibitem [{\citenamefont {Black}(2001)}]{black_introduction_2001}%
  \BibitemOpen
  \bibfield  {author} {\bibinfo {author} {\bibfnamefont {E.~D.}\ \bibnamefont
  {Black}},\ }\bibfield  {title} {\enquote {\bibinfo {title} {An introduction
  to {Pound}–{Drever}–{Hall} laser frequency stabilization},}\ }\href
  {https://doi.org/10.1119/1.1286663} {\bibfield  {journal} {\bibinfo
  {journal} {American Journal of Physics}\ }\textbf {\bibinfo {volume} {69}},\
  \bibinfo {pages} {79--87} (\bibinfo {year} {2001})}\BibitemShut {NoStop}%
\bibitem [{\citenamefont {Lemmerz}\ \emph {et~al.}(2017)\citenamefont
  {Lemmerz}, \citenamefont {Lux}, \citenamefont {Reitebuch}, \citenamefont
  {Witschas},\ and\ \citenamefont {Wührer}}]{lemmerz_frequency_2017}%
  \BibitemOpen
  \bibfield  {author} {\bibinfo {author} {\bibfnamefont {C.}~\bibnamefont
  {Lemmerz}}, \bibinfo {author} {\bibfnamefont {O.}~\bibnamefont {Lux}},
  \bibinfo {author} {\bibfnamefont {O.}~\bibnamefont {Reitebuch}}, \bibinfo
  {author} {\bibfnamefont {B.}~\bibnamefont {Witschas}},\ and\ \bibinfo
  {author} {\bibfnamefont {C.}~\bibnamefont {Wührer}},\ }\bibfield  {title}
  {\enquote {\bibinfo {title} {Frequency and timing stability of an airborne
  injection-seeded {Nd}:{YAG} laser system for direct-detection wind lidar},}\
  }\href {https://doi.org/10.1364/AO.56.009057} {\bibfield  {journal} {\bibinfo
   {journal} {Applied Optics}\ }\textbf {\bibinfo {volume} {56}},\ \bibinfo
  {pages} {9057} (\bibinfo {year} {2017})}\BibitemShut {NoStop}%
\bibitem [{\citenamefont {Dai}\ \emph {et~al.}(2018)\citenamefont {Dai},
  \citenamefont {Wu}, \citenamefont {Shi}, \citenamefont {Deng}, \citenamefont
  {Ge}, \citenamefont {Weng},\ and\ \citenamefont
  {Lin}}]{dai_development_2018}%
  \BibitemOpen
  \bibfield  {author} {\bibinfo {author} {\bibfnamefont {S.-T.}\ \bibnamefont
  {Dai}}, \bibinfo {author} {\bibfnamefont {H.-C.}\ \bibnamefont {Wu}},
  \bibinfo {author} {\bibfnamefont {F.}~\bibnamefont {Shi}}, \bibinfo {author}
  {\bibfnamefont {J.}~\bibnamefont {Deng}}, \bibinfo {author} {\bibfnamefont
  {Y.}~\bibnamefont {Ge}}, \bibinfo {author} {\bibfnamefont {W.}~\bibnamefont
  {Weng}},\ and\ \bibinfo {author} {\bibfnamefont {W.-X.}\ \bibnamefont
  {Lin}},\ }\bibfield  {title} {\enquote {\bibinfo {title} {Development of an
  injection-seeded single-frequency laser by using the phase modulated
  technique},}\ }\href {https://doi.org/10.1088/1674-1056/27/5/054212}
  {\bibfield  {journal} {\bibinfo  {journal} {Chinese Physics B}\ }\textbf
  {\bibinfo {volume} {27}},\ \bibinfo {pages} {054212} (\bibinfo {year}
  {2018})}\BibitemShut {NoStop}%
\bibitem [{\citenamefont {Cao}, \citenamefont {Li},\ and\ \citenamefont
  {Liu}(2020)}]{cao_theoretical_2020}%
  \BibitemOpen
  \bibfield  {author} {\bibinfo {author} {\bibfnamefont {X.}~\bibnamefont
  {Cao}}, \bibinfo {author} {\bibfnamefont {P.}~\bibnamefont {Li}},\ and\
  \bibinfo {author} {\bibfnamefont {Q.}~\bibnamefont {Liu}},\ }\bibfield
  {title} {\enquote {\bibinfo {title} {Theoretical and experimental
  investigation of injection seeded {Nd}:{YAG} zigzag slab ring lasers},}\
  }\href {https://doi.org/10.1016/j.optlastec.2019.105912} {\bibfield
  {journal} {\bibinfo  {journal} {Optics \& Laser Technology}\ }\textbf
  {\bibinfo {volume} {123}},\ \bibinfo {pages} {105912} (\bibinfo {year}
  {2020})}\BibitemShut {NoStop}%
\end{thebibliography}
\end{document}